\documentclass[twocolumn,amsmath,amssymb,aps,prb]{revtex4-1}

\usepackage{graphicx}
\usepackage{dcolumn} 
\usepackage{bm} 
\usepackage{hyperref} 
\usepackage{xcolor}

\allowdisplaybreaks    
\usepackage{multirow}

\begin{document}

\title{Anomalous Friedel oscillations in a quasi-helical quantum dot}
\author{F. M. Gambetta$^{1,2}$, N. Traverso Ziani$^{3}$, S. Barbarino$^{4}$, F. Cavaliere$^{1,2}$, M. Sassetti$^{1,2}$}
\affiliation{\noindent$^{1}$Dipartimento di Fisica, Universit\`{a} di Genova, Via Dodecaneso 33, I-16146 Genova, Italy\\
\noindent$^{2}$ SPIN-CNR, Via Dodecaneso 33, I-16146 Genova, Italy\\
\noindent$^{3}$Institute for Theoretical Physics and Astrophysics, University of W\"urzburg, 97074 W\"urzburg, Germany \\
\noindent$^{4}$ NEST, Scuola Normale Superiore \& Istituto Nanoscienze-CNR, I-56126 Pisa, Italy \\
}
\date{\today}


\begin{abstract}
The charge and spin patterns of a quantum dot embedded into a spin-orbit coupled quantum wire subject to a magnetic field are investigated. A Luttinger liquid theory is developed, taking into account open boundaries and finite magnetic field. In the quasi-helical regime, when spin-orbit effects dominate over the Zeeman interaction, peculiar states develop at the Fermi surface of the dot. {\em Anomalous} Friedel oscillations with twice the expected wavelength develop in the wavefunction of collective excitations of such states, accompanied by peculiar spin patterns in their magnetization. Both effects are analyzed in detail and shown possible to be probed in transport experiments. The stability against electron interactions and magnetic field is investigated. We also discuss how signatures of such states survive in the total charge and spin densities.
\begin{description}
\item[71.10.Pm;	71.70.Ej; 73.21.La]
\end{description}
\end{abstract}

\maketitle


\section{Introduction}

\noindent When electrons are confined in a tight one-dimensional portion of space, namely in a one-dimensional quantum dot, marked oscillations occur in the charge density~\cite{Jauregui:1993,Fabrizio:1995}.\\
\noindent {\em Friedel oscillations}~\cite{Giuliani_Vignale:2004} develop due to the presence of confining barriers which break the translational invariance and induce backscattering at the edges. Such oscillations exist regardless of the interactions between electrons, and give rise to a number of peaks in the charge density proportional to~\footnote{Here we assume for simplicity that $N_0$ and $D$ are commensurate} $N_0/D$, where $N_0$ and $D$ are the total number of electrons and the degeneracy of single particle levels respectively~\cite{Giuliani_Vignale:2004,Fabrizio:1995}. As an example, for a one-dimensional quantum wire of spinful electrons one would expect $N_0/2$ peaks for even $N_0$.\\
\noindent On the other hand strong interactions among the particles lead to the formation of peculiar correlated states, dubbed Wigner molecules~\cite{Reimann:2002,Yannoules:2007,Jauregui:1993,Hausler:1993,Szafran:2004,Abedinpour:2007,Secchi:2009,Qian:2010,Astrakharchik:2011}, the finite-size counterpart of Wigner crystals~\cite{Giuliani_Vignale:2004}. Such a molecule is characterized by $N_0$ peaks in the electron density~\cite{Giuliani_Vignale:2004,Reimann:2002,Yannoules:2007}, regardless of $D$.\\
\noindent Typically, in all systems with a degeneracy $D>1$, Friedel and Wigner oscillations have different wavelength and can be thus discriminated by looking at the charge density. On the other hand, in systems with $D=1$ - such as for instance a spin-polarized one-dimensional electron liquid - Friedel and Wigner oscillations have the same wavelength and cannot be distinguished at the level of single-particle density~\cite{Gambetta:2014}.\\
\noindent Many of these effects have been investigated in one-dimensional semiconducting quantum wires~\cite{Yacoby:1996} and carbon nanotubes~\cite{Pecker:2013}, in which a quantum dot can be created by defects~\cite{Yacoby:1996}, suitably crafted tunneling barriers or even buckling a carbon nanotube~\cite{Postma:2001}.

\noindent Recently, novel peculiar one-dimensional systems have been created, the helical liquids, occurring for example at the edges of topological insulators~\cite{Bernevig_Hughes_Zhang:2006,Wu:2006,Konig:2007} or in carbon nanotubes subject to an electric field~\cite{Klinovaja:2011}. In a helical liquid, electrons with opposite spin counter-propagate due to spin-momentum locking. In the presence of time-reversal (TR) symmetry, spin-momentum locking protects the chiral propagation of electrons and prevents elastic backscattering: indeed, non-magnetic barriers are not effective in confining the system. On the other hand, magnetic barriers can induce backscattering, leading to the formation of {\em spin density waves}.~\cite{Meng:2012} Two such barriers can create a quantum dot into the helical system, in which peculiar {\em spin textures} and spin ordering occurs and can be controlled by means of static or AC magnetic perturbations ~\cite{Timm:2012,Dolcetto:2013,Dolcetto:2013b,Dolcetto:2014}. However, such magnetic barriers do not give rise to charge oscillations in stark contrast with the previous case.\\

\noindent Spin-orbit coupled quantum wires subject to a magnetic field~\cite{Meng_Loss:2013,Streda_Seba:2003,Pershin:2004,Quay:2010} are also systems which have been lately investigated in depth. They exhibit a {\em quasi-helical} (Q-H) behavior different from the helical liquid discussed above. Indeed, the magnetic field breaks the TR symmetry mixing left- and right-movers and opens a Zeeman gap at zero momentum (see Fig.~\ref{spm:fig:Bands}). States in this gap display a peculiar character, due to the non-perfect spin-momentum locking. Since in these wires TR is broken by the magnetic field itself, a quantum dot can be created by means of usual non-magnetic barriers~\cite{Bjork:2005,Fasth:2007,Trif:2008}. In such a dot, charge oscillations will occur and due to the existing correlation between chirality and spin, peculiar spin textures are expected.\\

\noindent Such nanowires have been widely investigated in the last few years. When proximized with an s-wave superconductor, Majorana states occur at the boundaries of the wire~\cite{Klinovaja:2012}, which can be in principle investigated via STM transport~\cite{Liu:2013}. Also, an enhancement of the gap induced by electron interactions has been reported~\cite{Braunecker:2010}, with anisotropic spin properties~\cite{Meng_Loss:2013} and spin textures in the presence of magnetic impurities~\cite{Meng:2012}. Quasi-helical states also occur in wires with hyperfine coupling to the nuclear spins~\cite{Meng:2013}.\\
\noindent Such theoretical studies are based on a Luttinger model~\cite{Voit:1995,Von_Delft:1998,Giamarchi:2004} developed in the limit of vanishing applied magnetic field, typically employing periodic boundary conditions.

\noindent The task of this paper is to directly investigate the Q-H states which develop at finite magnetic field within the gap of a quantum dot and to assess how they affect the charge and spin densities.
\noindent To do so, we consider spin-orbit and a non-vanishing magnetic field on equal footing. Employing open-boundary conditions, states within the band gap are considered, for which a linearized spectrum and the corresponding wavefunctions for the single-particle problem are obtained. A Luttinger model with open boundaries is then developed, valid when the Fermi energy lies within the band gap, also in the presence of interactions among the electrons.\\
\noindent We study in details the charge distribution and the magnetization of states near the Fermi surface, introducing the concept of {\em collective excitation wavefunction} and {\em collective excitation magnetization}. They are the one-dimensional analogue of the quasiparticle wavefunction already introduced in literature and can be probed by means of STM transport experiments~\cite{Rontani:2005,Cavaliere:2009}. The stability of Q-H states against the intensity of the applied magnetic field and electron interactions is assessed.\\
\noindent We also investigate how the properties of such Q-H states reflect on those of the total charge and spin densities. Since these quantities also involve states below the Fermi surface one needs to go beyond the Luttinger theory. To do so, we have employed a Hubbard model solved by means of an exact diagonalization procedure in the absence of interactions and by a variational MPS algorithm in the interacting regime~\cite{White:1992,Schollwock:2011}.

\noindent Our main results are the following. When the spin-orbit effects dominate over the magnetic field, the Q-H states exhibit peculiar charge oscillations. In stark contrast to what one would naively expect for states with $D=1$, they {\em do not} exhibit $N_0$ peaks as for conventional Friedel oscillations, but rather $N_0/2$. We dub this an {\em anomalous Friedel oscillation} and show that it is intimately connected to the presence of evanescent states which form at the edges of the quantum dot within the band gap. In the Q-H regime, the length scale of these states becomes comparable to the dot size, resulting in the formation of anomalous Friedel oscillations. The latter are accompanied by {\em peculiar spin textures}: although the magnetization of the Q-H states precesses with $N_0$ peaks, strong modulations of the magnetization modulus occur near the dot edges, resulting in an effective doubling of the wavelength. Increasing the magnetic field or the interaction strength results into a progressive disappearance of the anomalous Friedel oscillations and ensuing spin textures with the dot edges being more stable with respect to its center. Signatures of the anomalous Friedel oscillations can be detected also in the total charge density, although the effects in this quantity are much less striking.

\noindent The paper is organized as follows. In Section~\ref{sec:the_model} we introduce the model and construct the Luttinger liquid theory in the presence of open boundary conditions. In Section~\ref{sec:results} we evaluate the collective excitation wavefunction and magnetization, describing the anomalous Friedel oscillations and the peculiar spin textures. We also discuss their stability against the magnetic field and electron interactions. Finally, we study the charge and spin densities. Section~\ref{sec:conclusions} contains the conclusions.


\section{The Model}
\label{sec:the_model}
\subsection{Single-particle problem}
\begin{figure}[h]
\centering
\includegraphics[width=0.7\columnwidth]{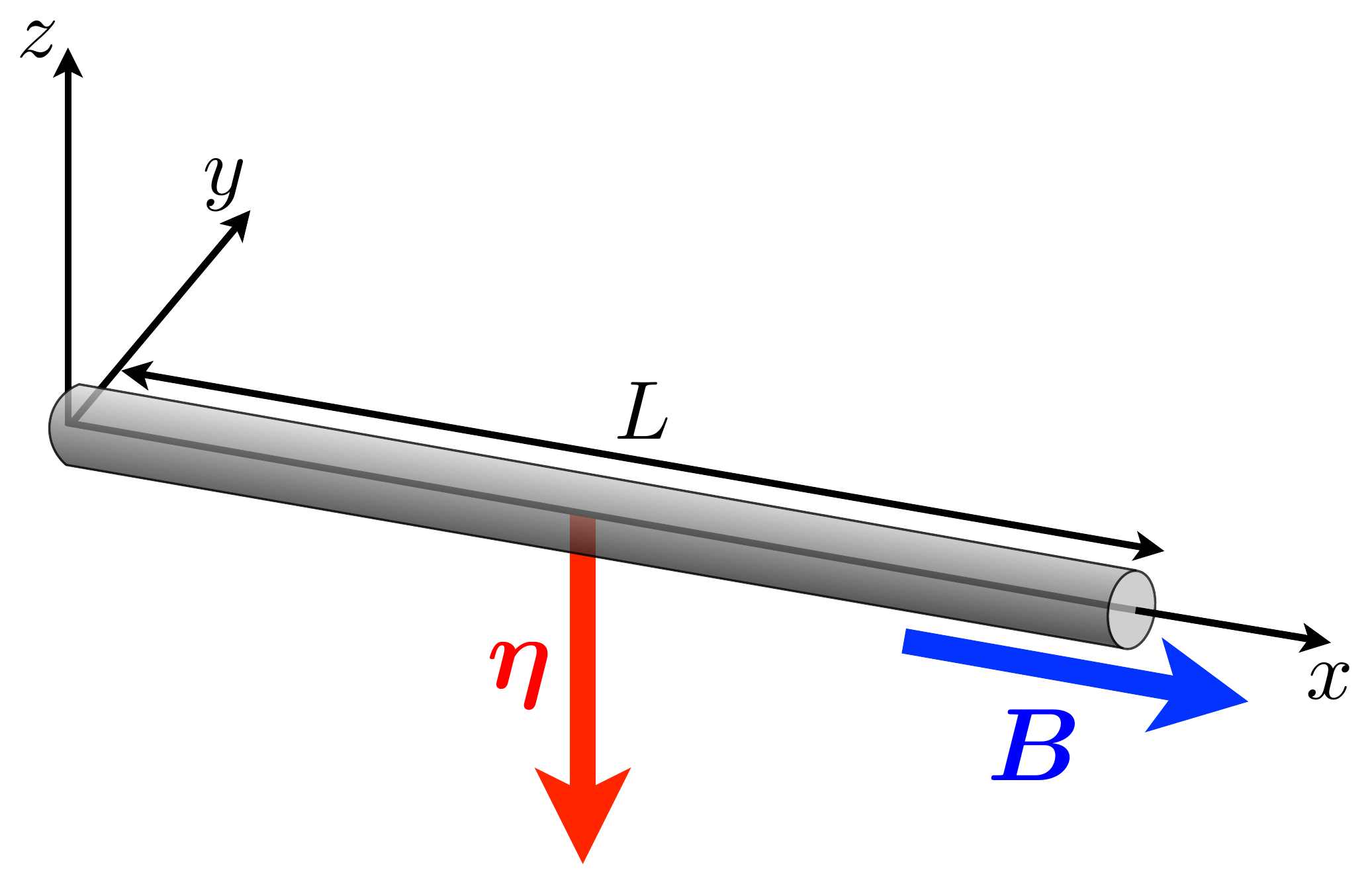}
\caption{Representation of the quantum dot of length $ L $ with intrinsic spin-orbit interaction, characterized by the vector $ \bm{\eta} $, subject to an external magnetic field $ \bm{B} $, parallel to the wire axis and orthogonal to the spin-orbit interaction.}
\label{spm:fig:Wire}
\end{figure}

%
%
We consider a quantum dot of length $L$, with a Rashba type spin-orbit interaction $ \bm{\eta}=-\eta\bm{u}_{z} $ ($ \eta>0 $) and subject to an external magnetic field $\bm{B}=B\bm{u}_{x}$ ($ B>0 $), with $ \bm{u}_{x} $ and $ \bm{u}_{z} $ axis unit vectors  (Fig.~\ref{spm:fig:Wire}).
Its single-electron Schr\"odinger equation ($ \hbar=1 $) is 
\begin{equation}
\left(-\frac{\partial_x^{2}}{2m^{*}}+i\eta\sigma_{z}\partial_{x}+\frac{1}{2}g^{*}\mu_{B}B\sigma_{x}\right)\Psi(x)=E\Psi(x).
\label{spm:eq:sp-aScheq}
\end{equation}
Here $\mu_{B} $ is the Bohr magneton, $ g^{*} $ and $ m^{*} $ are the effective gyromagnetic factor and the band mass of the electrons respectively, $ \sigma_{i} $ are the Pauli matrices and $ \Psi(x)=\left(\psi_{\uparrow}(x),\psi_{\downarrow}(x)\right)^{T} $ is the spinor eigenfunction satisfying open boundary conditions (OBC): $ \Psi(0)=\Psi(L)=0 $. To describe the system the following dimensionless parameters are useful
\begin{equation}
\alpha=\frac{\eta}{E_{0}L},\qquad\beta=\frac{g^*\mu_BB}{2E_0},\qquad\varepsilon=\frac{E}{E_0},
\label{spm:qe:parameters}
\end{equation}
where $ E_{0}=(2m^{*}L^{2})^{-1} $. In the rest of the paper all the energies will be written in units of $ E_0 $.
%
%
The spectrum obtained from Eq.~\eqref{spm:eq:sp-aScheq} consists of the two bands
\begin{equation}
\varepsilon_{\pm}(k)=L^2 k^2\pm\sqrt{\beta^2+\alpha^2 L^2 k^2}, 
\label{spm:eq:bands}
\end{equation}
where $ k\in\{k_{1,n}\} $, with $ \{k_{1,n}\}  $ a set of discrete wavevectors determined by the OBC, to be specified later. 
The magnetic field opens a gap $ \Delta=2\beta $ at $ k=0 $. The parameter
\begin{equation}
\delta = \frac{\beta}{\alpha^{2}}
\end{equation}
identifies two opposite regimes: the Q-H one for $ \delta < 1/2 $,~\cite{Streda_Seba:2003,Braunecker:2010,Meng_Loss:2013} dominated by spin-orbit, and the one for $ \delta > 1/2 $, where conversely the external magnetic field is prevalent and the system begins to polarize.
\begin{figure}
\centering
\includegraphics[width=0.9\columnwidth]{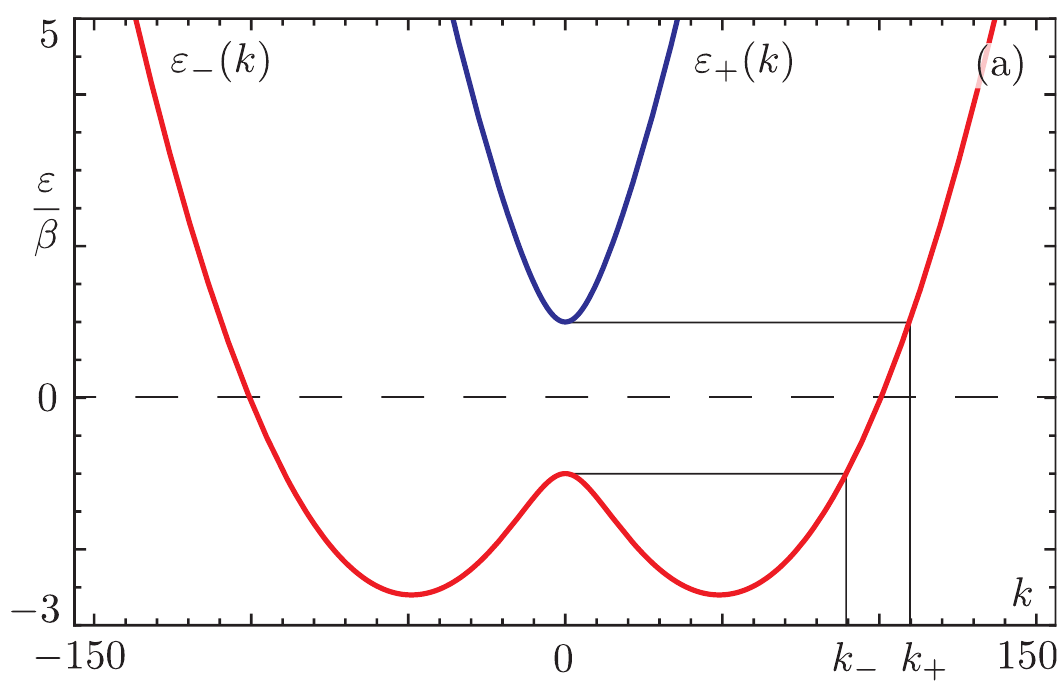}
\caption{Band structure of the quantum dot in the Q-H regime with $ k $ in units $ L^{-1} $. Parameters: $ \alpha=10^2,\,\beta=10^3 $ ($ \delta=0.1 $). }
\label{spm:fig:Bands}
\end{figure}
When $ \delta < 1/2 $ the band $ \varepsilon_{-}(k) $ has a maximum at $ k=0 $. The wavevectors inside the gap range from $ k_{-} $ to $ k_{+} $, with $ k_{\pm}=L^{-1}\sqrt{\alpha^2\pm2\beta} $ (see Fig.~\ref{spm:fig:Bands}). On the other hand, when $ \delta>1/2 $ the two minima vanish and the band $ \varepsilon_{-}(k) $ exhibits a single minimum in $ k=0 $ (not shown); in this case, $ k_{-}=0 $. Equation~\eqref{spm:eq:bands} can be formally solved for $ k $ as a function of $ \varepsilon $ with the result
\begin{equation}
k_{1,3}(\varepsilon)=\frac{1}{L}\sqrt{\frac{2\varepsilon+\alpha^2\pm\sqrt{\alpha^4+4\alpha^2\varepsilon+4\beta}}{2}},
\label{spm:eq:k1k3}
\end{equation}
where the $ + $ and $ - $ signs refer to $ k_1 $ and $ k_3 $ respectively. 

In this paper we are interested in studying energies inside the gap, i.e. $ |\varepsilon|<\beta $, where $ k_{1}L\in \mathbb{R}_{+} $ and $ k_{3}L $ is a purely imaginary number. It is then useful to rewrite it as $ k_3=i\kappa_3$, with  $\kappa_3L\in\mathbb{R}_{+}$.
Thus, the most general single-particle spinor wavefunction satisfying Eq.~\eqref{spm:eq:sp-aScheq} for $ |\varepsilon|<\beta $ has components
\begin{subequations}
\begin{align}
\psi_{\uparrow}(x)&\!=\!\frac{1}{\sqrt{2L}}\!\left(c_1e^{ik_1x}\!+\!c_2e^{-ik_1x}\!+\!c_3e^{-\kappa_3x}\!+\!c_4e^{\kappa_3x}\right)\!,\\
\psi_{\downarrow}(x)&\!=\!\frac{1}{\sqrt{2L}}\!\left(d_1e^{ik_1x}\!+\!d_2e^{-ik_1x}\!+\!d_3e^{-\kappa_3x}\!+\!d_4e^{\kappa_3x}\right)\!,
\end{align}
\label{spm:eq:general_psi_kappa}
\end{subequations}
with
\begin{subequations}
\begin{align}
d_{1}&=\frac{\varepsilon-L^2k_{1}^2+\alpha L k_{1}}{\beta}c_{1}\,,\\	
d_{2}&=\frac{\varepsilon-L^2k_{1}^2-L\alpha k_{1}}{\beta}c_{2}\,,\\	
d_{3}&=\frac{\varepsilon+L^2\kappa_{3}^2+i\alpha L\kappa_{3}}{\beta}c_{3}\,,\\	
d_{4}&=\frac{\varepsilon+L^2\kappa_{3}^2-i\alpha L\kappa_{3}}{\beta}c_{4}\,,
\end{align}
\label{spm:eq:d}
\end{subequations}
obtained from Eq.~\eqref{spm:eq:sp-aScheq}.


Imposing OBC, from Eqns.~\eqref{spm:eq:general_psi_kappa} and~\eqref{spm:eq:d} one gets a system of four linear equations that can be written as
\begin{equation}
\bm{M}\cdot\bm{c}=\bm{0},
\label{spm:eq:csystem}
\end{equation}
with $ \bm{M} $ a $ 4\times4 $ matrix of coefficients (not reported) and $ \bm{c}=(c_1,c_2,c_3,c_4)^{T} $. Imposing the condition $ \mathrm{Det}(\bm{M})=0 $ we obtain the secular equation
\begin{multline}
\alpha^2\sqrt{\beta^2-\varepsilon^2}\left[\cos(k_{1}L)\cosh(\kappa_{3}L)-1\right]=\\
(\alpha^2\varepsilon+2\beta^2)\sin(k_{1}L)\sinh(\kappa_{3}L),
\label{spm:eq:secular}
\end{multline}
which corresponds to an implicit equation for $ \varepsilon $.\\
\noindent In the following we will consider the solution of this equation in the limit $ \kappa_3L\gg1 $. Indeed, as shown in Appendix~\ref{sec:secular_equation}, this corresponds to have a large number of states in the gap, i.e. $k_{+}-k_{-}\gg\pi/L $, a necessary requirement in order to construct a Luttinger theory for states within the gap (see Sec.~\ref{subsec:Luttinger}).
Employing this condition and concentrating on the states near the center of the gap ($ |\varepsilon|\ll\beta $) one obtains analytical expressions for the linearized spectrum $\varepsilon_{k_{1,n}} $ and for the discretized wavevectors  $ k_{1,n} $ and $ \kappa_{3,n} $ (with $ n\in\mathbb{Z} $), see Appendix~\ref{sec:secular_equation}. The main results are:

\begin{align}
\varepsilon_{k_{1,n}}&=v_{0}\left[k_{1,n}-k_{1}^{(0)}\right], \label{spm:eq:en}\\
k_{1,n}&=\frac{\pi n}{L}+\frac{\pi}{2L}\gamma, \label{spm:eq:k1n}
\end{align}
where $ k_{1}^{(0)}\equiv k_{1}(\varepsilon=0)=\alpha L^{-1}[(\sqrt{1+4\delta^2}+1)/2]^{1/2} $ is the reference wavevector,
\begin{subequations}
\begin{gather}
v_{0}=\frac{\alpha^{2}\sqrt{1+4\delta^{2}}}{k_{1}^{(0)}}\label{spm:eq:v_0}\\
\intertext{the corresponding velocity and}
\gamma=1-\frac{2}{\pi}\arctan(2\delta). \label{spm:eq:gamma}
\end{gather}
\end{subequations}
We choose as a reference point $ n\approx n_{0} $, with $ \varepsilon_{n_{0}}\approx 0 $ the closest level to $ \varepsilon=0 $. As discussed in Appendix~\ref{sec:secular_equation}, one can always approximate $ \kappa_{3,n} $ as
\begin{equation}
\kappa_{3,n}\approx\kappa_3^{(0)}=\kappa_3(\varepsilon\!=\!0)=\frac{\alpha}{L}\!\left(\!\frac{\sqrt{1\!+\!4\delta^2}\!-\!1}{2}\!\right)^{\!1/2}\!.
\label{spm:eq:kappa3}
\end{equation}


The coefficients $ c_1,...,c_4 $ can be written as a function of $ c_1 $ only via the system in Eq.~\eqref{spm:eq:csystem}. For $ |\varepsilon|\ll\beta $ and $ \kappa_{3}^{(0)}L\gg1 $ they are
\begin{subequations}
\begin{align}
c_{2}\!&=\!\left\{\!\frac{1}{\chi}\!\left[\!\sqrt{2(\chi\!+\!1)}\!-\!1\right]\!-\!1-\!\frac{i}{\chi}\frac{\sqrt{\chi\!+\!1}\!-\!\sqrt{2}}{\sqrt{\chi\!-\!1}}\right\}c_{1},\\
c_{3}&=\frac{1}{\chi}\left[1-\sqrt{2(\chi+1)}+i\frac{\sqrt{\chi+1}-\sqrt{2}}{\sqrt{\chi-1}}\right]c_{1},\\
c_{4}&=(-1)^{n+1}\frac{e^{-\kappa_{3}^{(0)}L}}{\chi}\left[\frac{\sqrt{2}\chi-\sqrt{\chi+1}}{\sqrt{\chi-1}}+i\right]c_{1},
\end{align}
\end{subequations}
where $ \chi=\sqrt{1+4\delta^2} $. Coefficients $ d_{1},...,d_{4} $ are then obtained from Eq.~\eqref{spm:eq:d} and $ c_1\in\mathbb{R} $  is numerically determined from the normalization of the spinor. 

The above relations are valid \emph{in the whole range of $ \delta $}. However, the polarized regime ($ \delta>1/2 $), which can also be discussed in the framework of a spinless Luttinger liquid~\cite{Gambetta:2014}, is well known and not particularly novel. Thus, in this paper we will focus mostly on the Q-H regime, $ \delta<1/2 $. We emphasize here that even in the regime $ \delta\ll1/2 $, which is the most interesting one, we continue to assume the presence of a sufficiently large number of states in the gap with a finite magnetic field. This is possible since $ \alpha $ and $ \beta $ are independent parameters so that one can always tune $ \delta=\beta/\alpha^2\ll1/2 $ still satisfying the condition $ \kappa_3^{(0)}L=\beta/\alpha\gg1 $, see Appendix~\ref{sec:secular_equation}. In addition, in the Q-H regime the ratio between the number of states in the gap, $ \mathcal{N}_{g} $, compared to the total number of states in the Fermi sea, $ \mathcal{N}_{t} $, is $ \mathcal{N}_{g}/\mathcal{N}_{t} \propto \delta$. Thus in the Q-H regime the Fermi surface properties of the system are well described by our theory, conversely the physics of the whole Fermi sea is not necessarily captured. 

For future convenience analytical expression of the coefficients up to first order in $ \delta $ are quoted here:
 \begin{subequations}
 \begin{align}
 c_{1}&=1,\\
 c_{2}&=-i\frac{\delta}{2},\\
 c_{3}&=-1+i\frac{\delta}{2},\\
 c_{4}&=(-1)^{n+1}e^{-\kappa_{3}^{(0)}L}\left(i+\frac{3}{2}\delta\right),
 \end{align}
 \label{spm:eq:coeffc}
 \end{subequations}
 \begin{subequations}
 \begin{align}
 d_{1}&=-\frac{\delta}{2},\\
 d_{2}&=i+2\delta,\\
 d_{3}&=-\left(i+\frac{3}{2}\delta\right),\\
 d_{4}&=(-1)^{n}e^{-\kappa_{3}^{(0)}L}\left(-1+i\frac{\delta}{2}\right).
 \end{align}
  \label{spm:eq:coeffd}
 \end{subequations}
 
Substituting these coefficients in Eq.~\eqref{spm:eq:general_psi_kappa}, the Q-H nature of the states near the center of the gap is evident: in $ \psi_{\uparrow}(x) $ the leading term is the right-moving one, $ \propto e^{ik_1x} $, while in $ \psi_{\downarrow}(x) $ is the left-moving one, $ \propto e^{-ik_1x} $. In particular, at zero-order in $ \delta $ and ignoring the evanescent terms, the spinor $ \Psi(x) $ is analogous to that of a quantum spin Hall dot~\cite{Timm:2012,Dolcetto:2013}. 
 

\subsection{Luttinger liquid description}\label{subsec:Luttinger}

The linearized spectrum, obtained in the previous section (Eq.~\eqref{spm:eq:en}), allows to construct a LL theory with the Fermi energy $ \varepsilon_{F} $ lying near the center of the gap, i.e. $ |\varepsilon_{F}|\ll\beta $.
Let us introduce the fermionic field operator $ \hat{\Psi}(x)=(\hat{\psi}_{\uparrow}(x),\hat{\psi}_{\downarrow}(x))^{T} $, whose components are
\begin{multline}
\hat{\psi}_{\uparrow}(x)=\frac{1}{\sqrt{2L}}\sum_{k_{1,n}>0}^{\infty}\Big[c_1e^{ik_{1,n}x}+c_2e^{-ik_{1,n}x}\\
+c_3e^{-\kappa_3^{(0)}x}+\bar{c}_4(-1)^ne^{\kappa_3^{(0)}(x-L)}\Big]\hat{c}_{k_{1,n}},
\label{ll:eq:psi_up}
\end{multline}
with
\begin{equation}
\bar{c}_{4}=(-1)^{n}e^{\kappa_{3}^{(0)}L}c_{4},
\end{equation}
and a similar equation for $ \hat{\psi}_{\downarrow}(x) $ with $ c_{i}\rightarrow d_{i} $. Here $ \hat{c}_{k_{1,n}} $ is the fermionic operator annihilating the state $ \Psi_{k_{1,n}}(x)=(\psi_{\uparrow,k_{1,n}}(x),\psi_{\downarrow,k_{1,n}}(x))^{T} $.
In order to construct the Luttinger theory we introduce the right-mover field
\begin{equation}
\hat{\psi}_{R}(x)=\frac{1}{\sqrt{2L}}\sum_{k_{1,n}=-\infty}^{\infty}e^{ik_{1,n}x}\hat{c}_{k_{1,n}},
\label{ll:eq:psi_R}
\end{equation}
with wavevector extended from $ -\infty $ to $+\infty$.~\cite{Giamarchi:2004,Voit:1995,Von_Delft:1998} As a consequence, in the following all the operators must be redefined with respect to the vacuum state - i.e with no real electrons - $ |0 \rangle $ by means of the normal-ordering procedure~\cite{Von_Delft:1998} (denoted by $ :\quad: $). From Eq.~\eqref{spm:eq:k1n} one verifies that $ \hat{\psi}_{R}(x) $ satisfies the twisted boundary condition
\begin{equation}
\hat{\psi}_{R}(x+2L)=e^{i\pi\gamma}\hat{\psi}_{R}(x),
\end{equation}
where $ \gamma $ is given in Eq.~\eqref{spm:eq:gamma}.
As for a conventional LL~\cite{Fabrizio:1995}, OBC allow us to express the spinor field in terms of the only $ \hat{\psi}_{R}(x) $. We have
\begin{multline}
\hat{\psi}_{\uparrow}(x)=c_1\hat{\psi}_{R}(x)+c_2\hat{\psi}_{R}(-x)\\
+c_3e^{-\kappa_3^{(0)}x}\hat{\psi}_{R}(0)+\bar{c}_4e^{-i\frac{\pi}{2}\gamma}e^{\kappa_{3}^{(0)}(x-L)}\hat{\psi}_{R}(L),
\label{ll:eq:psi_upR}
\end{multline}
and similarly for $ \hat{\psi}_{\downarrow}(x) $ with $ c_i\rightarrow d_i $. In terms of $ \hat{\psi}_R(x) $ the non-interacting Hamiltonian becomes~\footnote{Here we have neglected terms $ \propto \gamma \text{ and } k^{(0)}_{1}$ which would lead to a contribution $ \propto \hat{N} $ in the Hamiltonian and thus to an irrelevant constant shift of the chemical potential of the dot.}
\begin{equation}
\hat{H}_0=v_{0}\int_{-L}^{L}\!:\hat{\psi}_{R}^{\dagger}(x)\left(\!-i\partial_x\right)\hat{\psi}_{R}(x)\!: dx.
\end{equation}
Following the standard procedure~\cite{Von_Delft:1998} $ \hat{\psi}_{R}(x) $ can be written via the bosonization formula
\begin{equation}
\hat{\psi}_{R}(x)=\frac{\hat{F}}{\sqrt{2\pi\Lambda}}e^{i\pi\frac{x}{L}(\hat{N}+\frac{1}{2}\gamma)}e^{i\hat{\Phi}(x)},
\end{equation} 
with $ \hat{F} $ the Klein factor and $ \hat{N}=\sum_{k_{1,n}}\!:\hat{c}_{k_{1,n}}^{\dagger}\hat{c}_{k_{1,n}}\!:$ the (normal-ordered) particle number operator. Here $ \Lambda $ is the cut-off length, set as $ \Lambda=L/\pi N_{0} $, with $ N_{0} $ the total number of electrons in the dot, and $ \hat{\Phi}(x) $ the bosonic field 
\begin{equation}
\hat{\Phi}(x)=\sum_{q>0}\sqrt{\frac{\pi}{Lq}}e^{iqx-\Lambda q/2}\hat{b}_{q}+\mathrm{h.c.}\, .
\label{ll:eq:bosonic_field}
\end{equation}
Here, $ q=\pi n_{q} / L $ ($ n_q $ a positive integer number) and $ \hat{b}^{\dagger}_q $, $ \hat{b}_q $ are bosonic creation and annihilation operators. 


The non-interacting Hamiltonian $ \hat{H}_0 $ can be bosonized as~\cite{Von_Delft:1998,Voit:1995,Giamarchi:2004} 
\begin{equation}
\hat{H}_{0}=v_{0}\sum_{q>0}q\hat{b}^{\dagger}_{q}\hat{b}_{q}+\frac{\pi v_{0}}{2L} \hat{N}^{2}.
\end{equation}


Let us now introduce the electron-electron interaction. As shown in Appendix~\ref{sec:InteractingH}, in the limit $ \kappa_{3}^{(0)}L \gg1$, the interacting Hamiltonian has the following form
\begin{equation}
\hat{H}_{\mathrm{int}}\!=\!\frac{V(0)}{2}\!\int_{-L}^{L}\!:\left[\hat{\rho}_{R}(x)\hat{\rho}_{R}(x) \!+\!\hat{\rho}_{R}(x)\hat{\rho}_{R}(-x)\right]\!: dx,
\label{ll:eq:H_int_reduced}
\end{equation}
where $ V(0) $ is the zero mode of the Fourier transform of a short range two-particle interaction and
\begin{equation}
\hat{\rho}_{R}(x)=:\hat{\psi}_{R}^{\dagger}(x)\hat{\psi}_{R}(x)\!:=\frac{\hat{N}}{2L}+\frac{\partial_{x}\hat{\Phi}(x)}{2\pi}
\label{ll:eq:rhoR}
\end{equation}
is the normal-ordered density of right-moving electrons. From Eq.~\eqref{ll:eq:rhoR} and making use of a Bogoliubov transformation, the total Hamiltonian $ \hat{H}=\hat{H}_{0}+\hat{H}_{\mathrm{int}} $ can be written in the diagonal form~\cite{Fabrizio:1995} 
\begin{equation}
\hat{H}=v\sum_{q>0}q\hat{d}_{q}^{\dagger}\hat{d}_{q}+\frac{\pi v_{N}}{2L}\hat{N}^{2},
\label{ll:eq:H}
\end{equation}
where $ \hat{d}^\dagger_q $, $ \hat{d}_q $ are the new bosonic creation and annihilation operators and $ v=v_{0}/g $, $ v_{N}=v_{0}/g^{2} $ are the velocities of bosonic and zero mode respectively. Here, $ v_{0} $ is the Fermi velocity introduced in Eq.~\eqref{spm:eq:v_0} and $ g=[1+V(0)/\pi v_0]^{-1/2} $ is the Luttinger parameter describing the intensity of the electron-electron interaction, with $ g<1 $ for repulsive interactions and $ g=1 $ for non-interacting electrons~\cite{Giamarchi:2004,Voit:1995}. In terms of the new bosonic operators $ \hat{d}_{q}^{\dagger} $ and $ \hat{d}_{q} $, the bosonic field in Eq.~\eqref{ll:eq:bosonic_field} becomes
\begin{equation}
\hat{\Phi}(x)\!\!=\!\!\frac{1}{\sqrt{g}}\!\sum_{q>0}\!\frac{e^{-\frac{\Lambda q}{2}}}{\sqrt{n_{q}}}\!\!\left\{\!\left[ \cos(qx)\!-\!ig\sin(qx) \right]\!\hat{d}_{q}^{\dagger} \!+ \!\mathrm{h.c.}\! \right\}\!.\\
\end{equation}

To be consistent with the linearization assumptions, it should be noted that the electron-electron interaction must be smaller than half the width of the band-gap.  This requirement is satisfied when $ g_{0}(\alpha,\delta)< g\leq1 $, with
\begin{equation}
g_{0}(\alpha,\delta)=\left[1+\frac{\alpha\delta}{2\sqrt{2}\pi}\sqrt{\frac{1+\sqrt{1+4\delta^2}}{1+4\delta^2}}\right]^{-\frac{1}{2}}.
\label{ll:eq:g_min}
\end{equation}


\section{Results}\label{sec:results}
Let us now discuss the charge and spin properties of the dot. Although we stress that our model is general and allows to explore the whole range of $\delta$, we will focus on the Q-H regime ($\delta< 1/2$), where the most striking features occur. As already noted, Q-H states lie within the gap. We will consider a dot filling such that the Fermi surface lies near the center of the band gap. In Sec.~\ref{sec:qpwf} we will concentrate on characterizing the Q-H states. In Sec.~\ref{sec:dens} we will discuss the total charge and spin densities, which involve the whole Fermi sea, employing a numerical approach based on the Hubbard model.
\subsection{Collective excitations wavefunction and magnetization}
\label{sec:qpwf}
Among the most powerful tools to investigate the properties of the states at the Fermi surface are the {\em collective excitation wavefunction} (CEWF)
\begin{equation}
\varphi(x)=\sum_{\sigma=\uparrow,\downarrow}\left|\langle N_0|\hat{\psi}^{\dagger}_{\sigma}(x)|N_0-1\rangle\right|^2\label{eq:qpwf}\, ,
\end{equation}
with $|N_0\rangle$ the ground state with $N_0$ electrons, and the {\em collective excitation magnetization} (CEM) $\mathbf{\Sigma}(x)=\left(\Sigma_{x}(x),\Sigma_{y}(x)\right)^{T}$ with
\begin{eqnarray}
\Sigma_{x}(x)&=&\sum_{p=\pm 1}\frac{p}{2}\left|\langle N_0|\hat{\psi}^{\dagger}_{\uparrow}(x)-p\hat{\psi}^{\dagger}_{\downarrow}(x)|N_0-1\rangle\right|^2 ,\ \label{eq:qpsx}\\
\Sigma_{y}(x)&=&\sum_{p=\pm 1}\frac{p}{2}\left|\langle N_0|\hat{\psi}^{\dagger}_{\uparrow}(x)+ip\hat{\psi}^{\dagger}_{\downarrow}(x)|N_0-1\rangle\right|^2,\ \label{eq:qpsy}
\end{eqnarray}
while the $z$ component vanishes identically for simmetry reasons, $\Sigma_{z}(x)\equiv 0$. The CEWF and $x$ component of CEM are even functions w.r.t. the dot center, i.e. $\varphi(x-L/2)=\varphi(x+L/2)$, $\Sigma_{x}(x-L/2)=\Sigma_{x}(x+L/2)$, while the $y$ component of the CEM is odd w.r.t. the dot center, i.e. $\Sigma_{y}(x-L/2)=-\Sigma_{y}(x+L/2)$, with $0\leq x\leq L/2$. One can directly verify that the amplitude of $\mathbf{\Sigma}(x)$ is given by
\begin{equation}
\Sigma(x)=\sqrt{\Sigma_{x}^{2}(x)+\Sigma_{y}^{2}(x)}\equiv\varphi(x)\label{eq:id0}\,.
\end{equation}
Equation~(\ref{eq:qpwf}) generalizes the notion of {\em quasiparticle wavefunction}, introduced in the context of circular quantum dots~\cite{Rontani:2005,Cavaliere:2009}, to the case of a Luttinger liquid. Such a quantity probes the probability density of states near the Fermi surface and, in the {\em non-interacting} regime $g=1$, reduces to
\begin{equation}
\varphi(x)=\varphi_{0}(x)\equiv \left|\Psi_{N_{0}}(x)\right|^2\, ,\label{eq:id1}
\end{equation}
where $\Psi_{N_{0}}(x)$ is the spinor wavefunction of the $N_{0}$--th level. Analogously, Eqns.~(\ref{eq:qpsx}, \ref{eq:qpsy}) describe the magnetization of states near the Fermi surface. For $g=1$ one has 
\begin{equation}
\Sigma_{\nu}(x)= \Sigma_{0}^{(\nu)}(x)\equiv\left(\Psi_{N_{0}}^{*}(x)\right)^{T}\hat{\sigma}_{\nu}\Psi_{N_{0}}(x)\, ,\label{eq:id2}
\end{equation}
with $\nu\in\{x,y\}$.\\ 
\noindent Both $\varphi(x)$ and $\mathbf{\Sigma}(x)$ can be extracted via tunneling experiments involving a magnetized STM tip. In particular, one has
\begin{eqnarray}
\varphi(x)&\propto&\sum_{p=\pm 1}\Gamma_{p}^{(\nu)}(x)\quad\forall\nu,\label{eq:ratetot}\\
\Sigma_{\nu}(x)&\propto&\sum_{p=\pm1}p\Gamma_{p}^{(\nu)}(x),\label{eq:ratesp}
\end{eqnarray}
where $\Gamma_{p}^{(\nu)}(x)$ is the tunneling rate for injecting electrons from a tip with spin component $p$ along the quantization axis $\nu$. See Appendix~\ref{sec:appstm} for details.\\

\noindent The Luttinger liquid theory developed here allows to analytically evaluate the CEWF and CEM for any value of $\delta$ also in the presence of interactions. Although such evaluation can be performed for any temperature, in the following we will focus on the most interesting regime $k_{B}T\ll\pi v_0/L$, where expectation values can be performed in the $T\to 0$ limit. Useful analytic expressions for $\delta\ll1/2$ can also be obtained using coefficients in Eqns.~(\ref{spm:eq:coeffc}, \ref{spm:eq:coeffd}). Up to first order in $\delta$ one has ($x<L/2$)
\begin{eqnarray}
\varphi(x)\!&=&\!\frac{K_g(0,0)\!-\!2K_g(0,x)\cos(k_{F}x)\!+\!K_g(x,x)}{\pi\Lambda}\nonumber\\
\!\!\!\!\!\!&+&\!\frac{\delta}{\pi\Lambda}\left[2K_g(0,x)\sin(k_{F}x)\!-\!K_g(x,x)\sin(2k_{F}x)\right],\label{eq:cewf}\\
\Sigma_{x}(x)\!&=&\!\frac{2K_g(0,x)\sin(k_{F}x)-K_g(x,x)\sin(2k_{F}x)}{\pi\Lambda}\nonumber\\
&+&\!\frac{\delta}{\pi\Lambda}\left\{-K_g(0,0)\!+\!2K_g(0,x)\cos(k_{F}x)\right.\nonumber\\
&+& \left.K_g(x,x)\left[1-2\cos(2k_{F}x)\right]\right\},\label{eq:cemx}\\
\Sigma_{y}(x)\!&=&\!\frac{K_g(0,0)\!-\!2K_g(0,x)\cos(k_{F}x)\!+\!K_g(x,x)\cos(2k_{F}x)}{\pi\Lambda}\nonumber\\
&+&\!\frac{2\delta}{\pi\Lambda}\left[2K_g(0,x)\sin(k_{F}x)\!-\!K_g(x,x)\sin(2k_{F}x)\right],\label{eq:cemy}
\end{eqnarray}
where we have neglected terms $\propto\exp\left[\kappa_{3}^{(0)}(x-L)\right]\ll 1$ for $x<L/2$. Furthermore,
\begin{equation}
K_g(x_1,x_2)=e^{-k_{3}^{(0)}[2x-(x_1+x_2)]}W_g(x_1)W_g(x_2)\, ,\label{eq:K}	
\end{equation} 
with
\begin{eqnarray}
W_g(x)&=&\left[1-2e^{-\pi\Lambda/L}\cos\left(\frac{2\pi x}{L}\right)+e^{-2\pi\Lambda/L}\right]^{\frac{1}{8}\left(\frac{1}{g}-g\right)}\nonumber\\
&&\times\left(1-e^{-\pi\Lambda/L}\right)^{\frac{1}{4}\left(\frac{1}{g}+g\right)}
\end{eqnarray}
and $k_{F}=\pi N_{0}/L + \pi \gamma/2L$ the Fermi momentum for $N_{0}$ electrons. The above expressions consist of non-oscillating terms $\propto K_{g}(0,0)$, terms $\propto K_{g}(0,x)$ oscillating with wavevector $k_{F}$ and terms $\propto K_{g}(x,x)$ oscillating with wavevector $2k_{F}$. Constant terms and terms oscillating with wavevector $k_{F}$ are enveloped by $\exp\left[-2k_{3}^{(0)}x\right]$ and $\exp\left[-k_{3}^{(0)}x\right]$ respectively. To lowest order in $\Lambda/L$, one has that $W_{g}(x\approx 0)\propto(\pi\Lambda/L)^{1/2g}$, while $W_{g}(x\approx L/2)\propto(\pi\Lambda/L)^{(g+1/g)/4}$. Furthermore, for $ g=1 $ one has $W_1(x)=\left(\pi\Lambda/L\right)^{1/2}$ $\forall x$.\\
\noindent In the non-interacting case, the above expressions reduce therefore to
\begin{eqnarray}
\varphi(x)&=&\frac{1}{L}\Big\{ 2e^{-\kappa_{3}^{(0)}x}\big[\cosh(\kappa_{3}^{(0)}x)-\cos(k_{F}x)\big]\nonumber\\
&+&\delta\big[2e^{-\kappa_{3}^{(0)}x}\sin(k_{F}x)-\sin(2k_{F}x)\big]\Big\}\, ,\label{eq:fmsq}\\
\Sigma_{x}(x)&=& \frac{1}{L}\Big\{ 2e^{-\kappa_{3}^{(0)}x}\sin(k_{F}x)-\sin(2k_{F}x)\nonumber\\
&-& 2\delta\big[\cos(2k_{F}x)-e^{-\kappa_{3}^{(0)}x}\cos(k_{F}x)\nonumber\\
&-&e^{-\kappa_{3}^{(0)}x}\sinh(\kappa_{3}^{(0)}x)\big]\Big\}\, ,\label{eq:fspx}\\	
\Sigma_{y}(x)&=& \frac{1}{L}\Big\{\cos(2k_{F}x)-2e^{-\kappa_{3}^{(0)}x}\cos(k_{F}x)+e^{-2\kappa_{3}^{(0)}x}\nonumber\\
&-& 2\delta\big[\sin(2k_{F}x)-2e^{-\kappa_{3}^{(0)}x}\sin(k_{F}x)\big]\Big\}\, ,\label{eq:fspy}
\end{eqnarray}
which coincide, as anticipated, with the quantities $\varphi_{0}(x)$ in Eq.~(\ref{eq:qpwf}) and $\Sigma_{0}^{(\nu)}(x)$ in Eqns.~(\ref{eq:qpsx}, \ref{eq:qpsy}) evaluated for $x<L/2$ to the first order in $\delta$, as can be directly verified employing the definition of the dot spinor wavefunction in Eq.~\eqref{spm:eq:general_psi_kappa} and neglecting terms proportional to $\exp\left[\kappa_3^{(0)}(x-L)\right]$ as discussed above.
\begin{figure}
\centering
\includegraphics[width=\columnwidth]{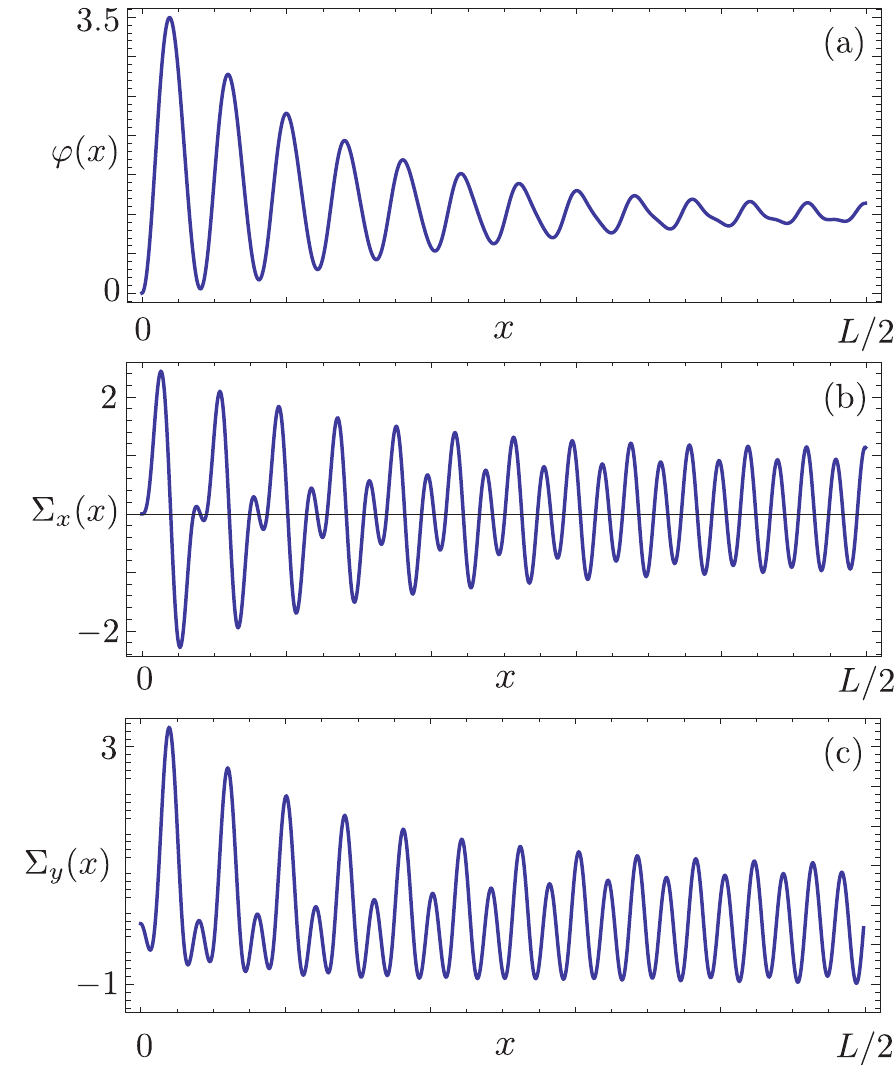}
\caption{Plot of the CEWF (a) and of the $x,y$ components of the CEM (b, c) for $N_0=49$ non-interacting electrons in the Q-H regime with $\delta=0.044$. Parameters here: $\alpha=150$, $\beta=1000$ and $g=1$.}
\label{fig:figr1}
\end{figure}

\noindent In the Q-H regime the CEWF and the CEM exhibit markedly enveloped oscillations as shown in Fig.~\ref{fig:figr1} for non-interacting electrons. We start considering $N_0=49$ and $\delta=0.044\ll 1$, deep in the Q-H regime.\\
\noindent The CEWF (panel (a)) displays oscillations with a wavevector $k_{F}$ and $N_{p}=25$ peaks (taking into account that $\varphi(x)$ is even w.r.t. the dot center). For any $N_0$ one has
\begin{equation}
N_p=\begin{cases}
N_0/2 & \text{if }N_0\text{ is even}\, ,\\
(N_0+1)/2 & \text{if }N_0\text{ is odd}\, .
\end{cases}	
\end{equation}
This behavior is in contrast with the expectations for the standard Friedel oscillations of a one-dimensional system with non-degenerate states~\cite{Fiete:2006,Fiete:2007,Gindikin_Sablikov:2002,Gindikin_Sablikov:2007,Gambetta:2014}, which instead would predict oscillations at wavevector $2k_{F}$, with $N_0$ peaks. We dub this an {\em anomalous Friedel oscillation}. The CEWF oscillations are more pronounced near the edges, while near the center $\varphi(x)$ is flatter.\\
\noindent On the other hand, the oscillations of $\Sigma_{\nu}(x)$ (panels (b, c)) display $N_0$ peaks corresponding to a wavevector $2k_{F}$, in agreement with the expectations for the standard Friedel oscillations, with $\Sigma_{x}(x)$ and $\Sigma_{y}(x)$ essentially out of phase by $\pi/2$ - suggesting a precessing pattern of the CEM which strongly resembles that of a helical system~\cite{Dolcetto:2013b,Dolcetto:2013}. However, CEM oscillations near the edges are far less regular than those in the center. Indeed, near the edges an alternating sequence of high and low peaks emerges.
\begin{figure}
\centering
\includegraphics[width=\columnwidth]{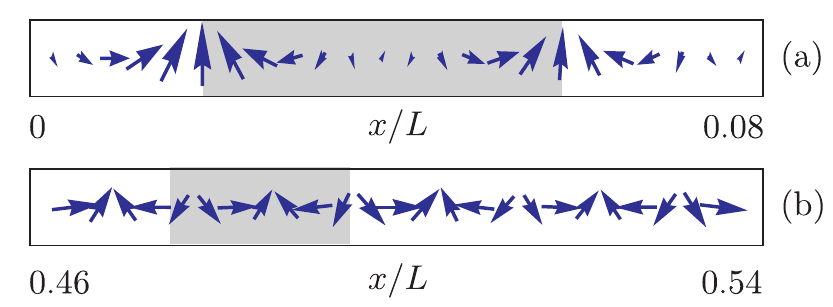}
\caption{Vector plot of $\bm{\Sigma}(x)$ (a) near the edge and (b) near the center of the dot for the Q-H case $\delta=0.044$. Gray shades denote the quasi-period of the magnetization pattern. Parameters as in Fig.~\ref{fig:figr1}.}
\label{fig:figr2}
\end{figure}
To better visualize the spin texture, Fig.~\ref{fig:figr2} shows the spatial pattern of the CEM vector in different dot regions. Near the edge (panel (a)), where the anomalous Friedel oscillations are stronger, the CEM vector length is strongly modulated and an effective {\em anomalous} oscillation of the magnetization, with wavevector $k_{F}$, emerges. This is a direct consequence of the anomalous Friedel oscillation. Indeed, the {\em amplitude} of $\mathbf{\Sigma}(x)$ is given by the charge oscillations in $\varphi(x)$, as shown in Eq.~\eqref{eq:id0} . On the other hand, in the center (panel (b)) the magnitude of the CEM is only slightly modulated, allowing to observe a more conventional precession with wavevector $2k_{F}$. Therefore, a mixed character of the spin pattern emerges, with different periodicities between the edges ($k_{F}$) and the dot center ($2k_{F}$).\\

\noindent The anomalous Friedel oscillations and the peculiar spin textures are a genuine hallmark of the Q-H regime. They are driven by the exponential edge terms ($\propto\exp[-\kappa_{3}^{(0)}x]$ for $ x<L/2 $ and $\propto\exp[\kappa_{3}^{(0)}(x-L)]$ for $ x>L/2 $), peculiar of states in the band gap, with  a typical length scale $\ell=1/\kappa_{3}^{(0)}$. For $\alpha=150$ and $\delta=0.044$  - see Fig.~(\ref{fig:figr1}) - one has $\kappa_{3}^{(0)}L\approx 6.5$ and thus $\ell\approx 0.15 L$. In such a situation, oscillations with wavevector $k_{F}$ fully emerge in the CEWF and the peculiar spin texture with different periodicities at the dot edges and center develops in the CEM.\\
\noindent To get a clearer picture of how this may occur, let us analyze the approximate expressions for $\varphi(x)$ and $\mathbf{\Sigma}(x)$ in Eq.~(\ref{eq:fmsq}-\ref{eq:fspy}). For $\delta\ll 1/2$, as in the case discussed so far, the leading term for the CEWF is the sole oscillation with wavevector $k_{F}$. On the other hand, the CEM components display both oscillations at $k_{F}$ and $2k_{F}$ - see Eqns.~(\ref{eq:fspx},\ref{eq:fspy}). The mixing between the two is controlled by the exponential term which damps oscillations with wavevector $k_{F}$ near the dot center. This mixing is responsible for the irregular, alternating pattern of high and low peaks in Figs.~\ref{fig:figr1}(b, c).
\begin{figure}
\centering
\includegraphics[width=\columnwidth]{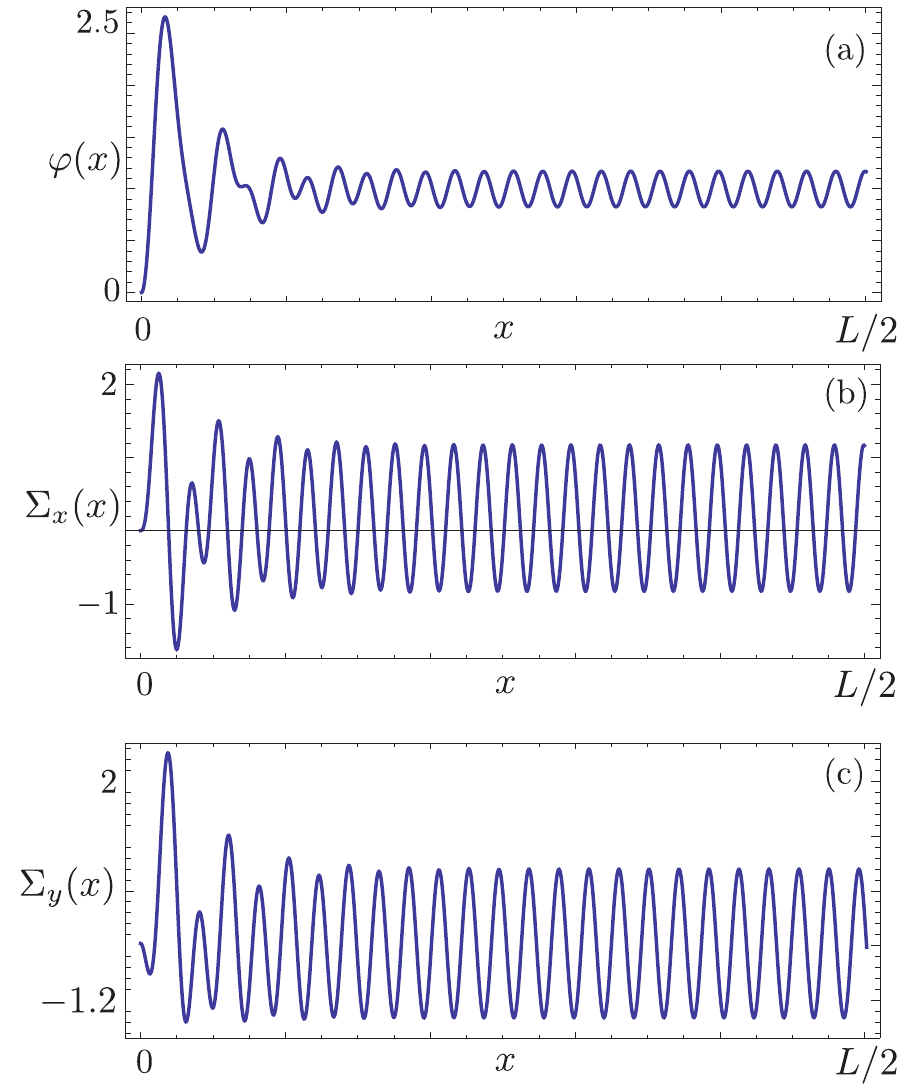}
\caption{Plot of the CEWF (a) and of the $x,y$ components of the CEM (b,c) for $N_0=49$ non-interacting electrons in the Q-H regime with $\delta=0.177$. Parameters here: $\alpha=150$, $\beta=4000$ and $g=1$.}
\label{fig:figr3}
\end{figure}
\begin{figure}
\centering
\includegraphics[width=\columnwidth]{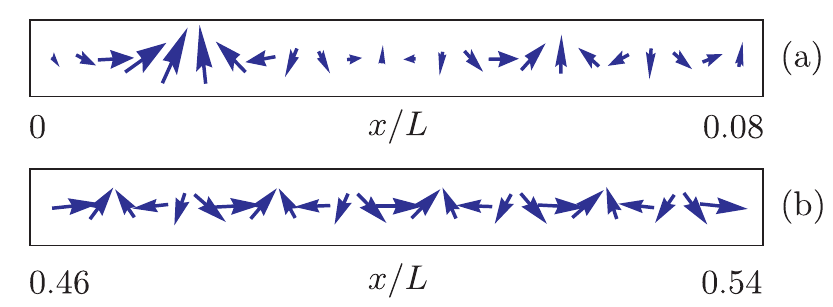}
\caption{Vector plot of $\bm{\Sigma}(x)$ (a) near the edge and (b) near the center of the dot for $\delta=0.177$. Parameters as in Fig.~\ref{fig:figr3}.}
\label{fig:figr4}
\end{figure}
Increasing $\delta$, two phenomena occur. Firstly, $\kappa_{3}^{(0)}$ increases and consequently $\ell$ shrinks, resulting in a suppression of the anomalous Friedel oscillations of the CEWF in the center, accompanied by a shrinkage of the region where the mixing between oscillations at $k_{F}$ and $2k_{F}$ occur in the CEM. In addition, corrections to the picture valid in the Q-H regime occur, signaled by the terms $\propto\delta$ in Eqns.~(\ref{eq:fmsq}-\ref{eq:fspy}). As a result, conventional Friedel oscillations with wavevector $2k_{F}$ emerge in the CEWF. They are expected to be stroger near the center, while anomalous Friedel oscillations still survive in a region with length of order $\ell$ near the edges.\\
\noindent Anomalous Friedel oscillations and peculiar CEM patterns are expected until $\ell\gtrsim L/N_{0}$, the latter being the typical wavelength of the conventional Friedel oscillations.\\
\noindent Our predictions are supported by Fig.~\ref{fig:figr3} and Fig.~\ref{fig:figr4}, which show the results for $\delta=0.177$ (with $\ell\approx 0.05L>L/N_0$). The CEWF displays a regular pattern of conventional Friedel oscillations in the center and only few anomalous Friedel oscillations survive near the edges. Also the CEM displays a far more regular behavior, with much less evident alternating high-low peaks near the dot edges. This also reflects in a more regular precession of the vector $ \bm{\Sigma}(x) $ in the latter region, as shown in Fig.~\ref{fig:figr4}(a).
\noindent Increasing $\delta>1/2$, the dot totally leaves the Q-H regime and becomes polarized. Here, the dot displays conventional Friedel oscillations while the spin begins to become polarized along the $x$ direction (not shown).~\cite{Gambetta:2014}\\

\noindent Let us now discuss interaction effects in the Q-H regime, with the aid of the approximate expressions in Eqns.~(\ref{eq:cewf}-\ref{eq:K}). For $g<1$ one can understand the fate of the anomalous Friedel oscillations inspecting the power-law scaling of $K_{g}(x_1,x_2)$. Near the dot center, for $x\approx L/2$, one has
\begin{subequations}
\begin{align}
K_g(0,0)&\propto\left(\frac{\pi\Lambda}{L}\right)^{\frac{1}{g}},\\ K_g(0,x)&\propto\left(\frac{\pi\Lambda}{L}\right)^{\frac{3}{4g}},\\ K_g(x,x)&\propto\left(\frac{\pi\Lambda}{L}\right)^{\frac{1}{2g}} .
\end{align}
\end{subequations}
Thus, although near the center both $\varphi(x)$ and $\mathbf{\Sigma}(x)$ vanish as interactions are increased, anomalous Friedel oscillations $\propto K_{g}(0,x)$ tend to zero with a faster power law than conventional Friedel oscillations $\propto K_{g}(x,x)$. On the other hand, near the edges $x\approx 0,L$ all three terms scale with the same power law $\propto (\pi\Lambda/L)^{1/2g}$. Consequently, as the interaction strength is increased, the region where the anomalous Friedel oscillations can be observed shrinks near the dot edges.
\begin{figure}
\centering
\includegraphics[width=\columnwidth]{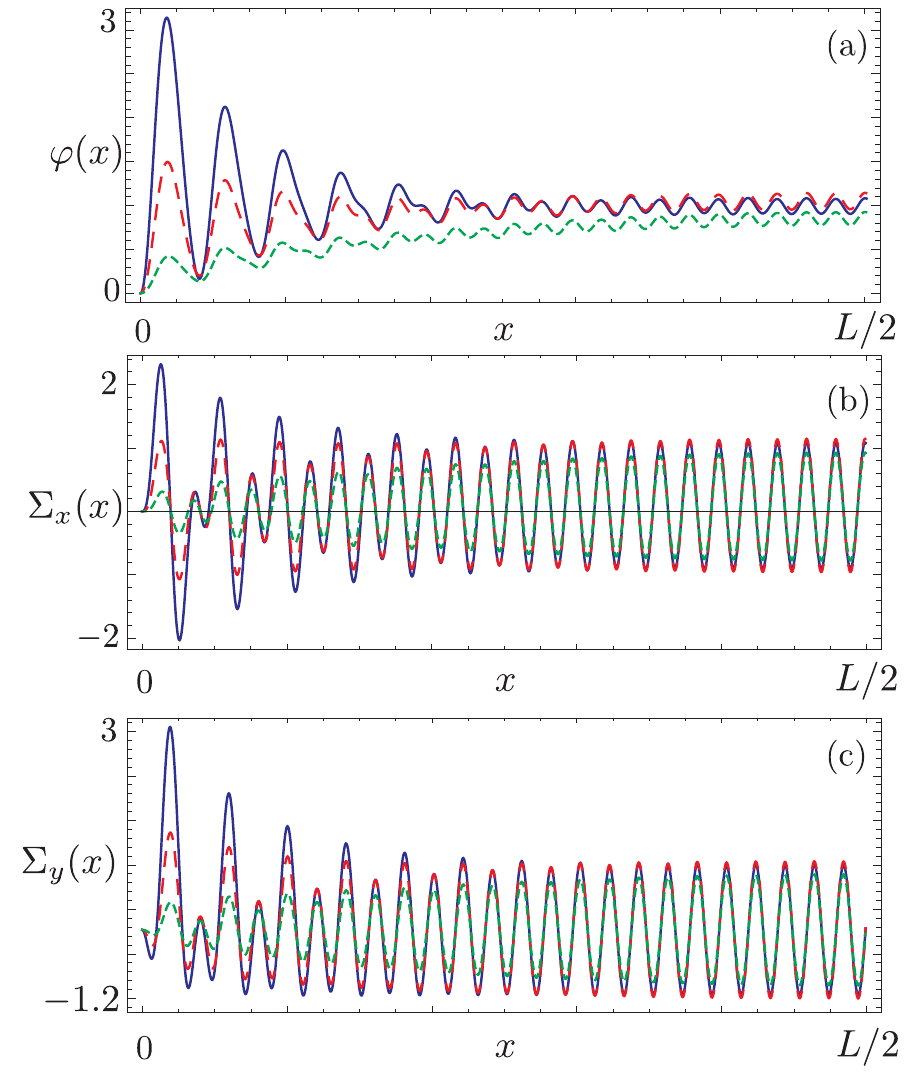}
\caption{Plot of the CEWF (a) and of the $x,y$ components of the CEM (b,c) for $N_0=49$, $\delta=0.177$ and different values of the interaction parameter: $g=1$ (blue, solid), $g=0.8$ (red, dashed) and $g=0.6$ (green, dotted). Parameters here: $\alpha=150$, $\beta=4000$.}
\label{fig:figr5}
\end{figure}
Our analysis is supported by the behavior of the CEWF and CEM for different values of $g$ and a fixed value of $\delta$ as depicted in Fig.~\ref{fig:figr5}. For a given interaction strength the behavior as a function of $\delta$ remains qualitatively similar to that for non-interacting electrons already discussed above.\\

\subsection{Charge and spin densities}
\label{sec:dens}
We now investigate the visibility of the phenomena observed previously at the Fermi surface in the Q-H regime in the total charge $\hat{\rho}(x)$ and spin densities $\hat{\mathbf{S}}(x)=\left(\hat{S}_{x}(x),\hat{S}_y(x)\right)^{T}$, which involve the whole Fermi sea. As already anticipated in Sec.~\ref{sec:the_model}, the ratio between the number of states in the gap and the total number of states in the Fermi sea scales as $\delta$, so that in the Q-H regime with $\delta<1/2$ a different approach should be used to evaluate the above quantities. We have employed a Hubbard model~\cite{Hubbard:1963,Essler:2005}: the dot has been discretized into $\mathcal{N}$ sites and corresponding fermionic operators $\hat{c}_{j,\sigma}$ are introduced. The Hamiltonian $\hat{H}_{\mathrm{Hub}}=\hat{H}_{\mathrm{TB}}+\hat{H}_{\mathrm{int}}$ is
\begin{eqnarray}
	\hat{H}_{\mathrm{TB}} \!&=& \!\left( \!-t \!-\!i \frac {\eta}{2} \right) \!\sum_{j=1}^{\mathcal{N}-1} \!  \hat{c}^\dagger_{j, \uparrow} \hat{c}_{j+1, \uparrow} \! + \!\left( \!-t \!+\! i  \frac {\eta}{2} \right) \!\sum_{j=1}^{\mathcal{N}-1} \! \hat{c}^\dagger_{j+1,\uparrow} \hat{c}_{j,\uparrow} \nonumber\\
	&+&\! \left( \!-t \!- \!i  \frac {\eta}{2} \right) \!\sum_{j=1}^{\mathcal{N}-1} \! \hat{c}^\dagger_{j, \downarrow}\hat{c}_{j+1, \downarrow}  \!+ \!\left( \!-t\! +\! i  \frac {\eta}{2} \right) \!\sum_{j=1}^{\mathcal{N}-1}  \!\hat{c}^\dagger_{j+1,\downarrow} \hat{c}_{j,\downarrow}  \nonumber\\
	&+&\frac{1}{2}g^{*}\mu_{B}B \sum_{j=1}^{\mathcal{N}}  \left( \hat{c}^\dagger_{j, \uparrow} \hat{c}_{j, \downarrow}+ \hat{c}^\dagger_{j, \downarrow}  \hat{c}_{j, \uparrow}  \right)\label{eq:h0hub}\, ,\\
\hat{H}_{\mathrm{int}} &=& U\sum_{j=1}^{\mathcal{N}}\hat{c}_{j,\uparrow}^{\dagger}\hat{c}_{j,\uparrow}\hat{c}_{j,\downarrow}^{\dagger}\hat{c}_{j,\downarrow}
\end{eqnarray}
Here, $t$ is the hopping amplitude between neighboring sites and $U>0$ the strength of the repulsive on-site electron interaction. We are interested into $\rho(x)=\langle\hat{\rho}(x)\rangle$ and $S_{\nu}(x)=\langle \hat{S}_{\nu}(x) \rangle$, where $\langle\ldots\rangle$ denotes the zero-temperature quantum average valid in the low-temperature limit $k_{B}T\ll\pi v_0/L$. Note that $S_{z}(x)=\langle \hat{S}_{z}(x)\rangle\equiv 0$ and that $\rho(x)$ and $S_{\nu}(x)$ share the same spatial symmetries of $\varphi(x)$ and $\Sigma_{\nu}(x)$ respectively.\\

\noindent Let us begin considering $N_0=48$ non-interacting electrons, which can be treated by means of an exact diagonalization.
\begin{figure}
\centering
\includegraphics[width=\columnwidth]{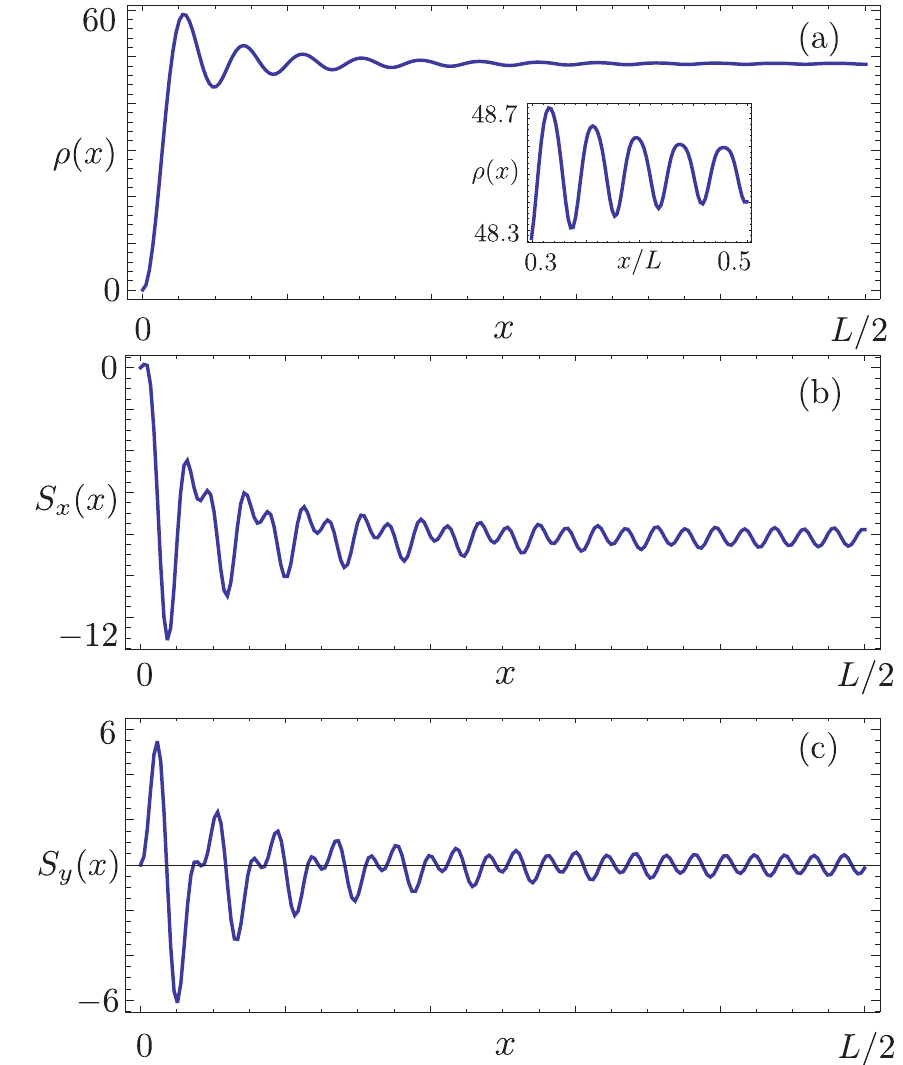}
\caption{Charge density (a) (units $L^{-1}$) and spin $x$ (b) and $y$ (c) components densities (units $L^{-1}$) for a quantum dot with $N_0=48$ non-interacting electrons in the Q-H regime $\delta=0.044$. The inset in (a) shows the weak density oscillations near the dot center. Parameters: $L=8.4\mu\mathrm{m}$, $\mathcal{N}=432$, $t=4.5\ \mathrm{meV}$, $U=0$, $\alpha=150$, $\beta=1000$.}
\label{fig:fig1h}
\end{figure}
\noindent Figure~\ref{fig:fig1h} shows the charge and spin densities for $\delta=0.044$, deep in the Q-H regime. The total charge density exhibits oscillations centered around the average value $N_{0}/L$ as expected with $N_{p}=24$ peaks, consistently with the anomalous Friedel oscillations of the CEWF. This number of peaks is also what one would naively expect from standard Friedel oscillations considering that the states below the gap have a parabolic spectrum with $D=2$. However, the influence of Q-H states at the Fermi surface gives rise to a much flatter density in the center and more pronounced oscillations near the edges. This confirms the picture discussed for the CEWF in the previous section.\\

\noindent Also the spin density $\mathbf{S}(x)$ displays signatures induced by the anomalous Friedel oscillations. Both $S_{x}(x)$ and $S_{y}(x)$ display $N_0$ peaks consistent with a $2k_{F}$ oscillation. However, an anomalous alternating pattern of high and low peaks is observed in both quantities near the dot edges, in strong analogy with the behavior of the CEM components. We note in passing that $S_{x}(x)$ oscillates around a non-zero reference level, connected to the partial polarization of the dot induced by the external magnetic field.
\begin{figure}
\centering
\includegraphics[width=\columnwidth]{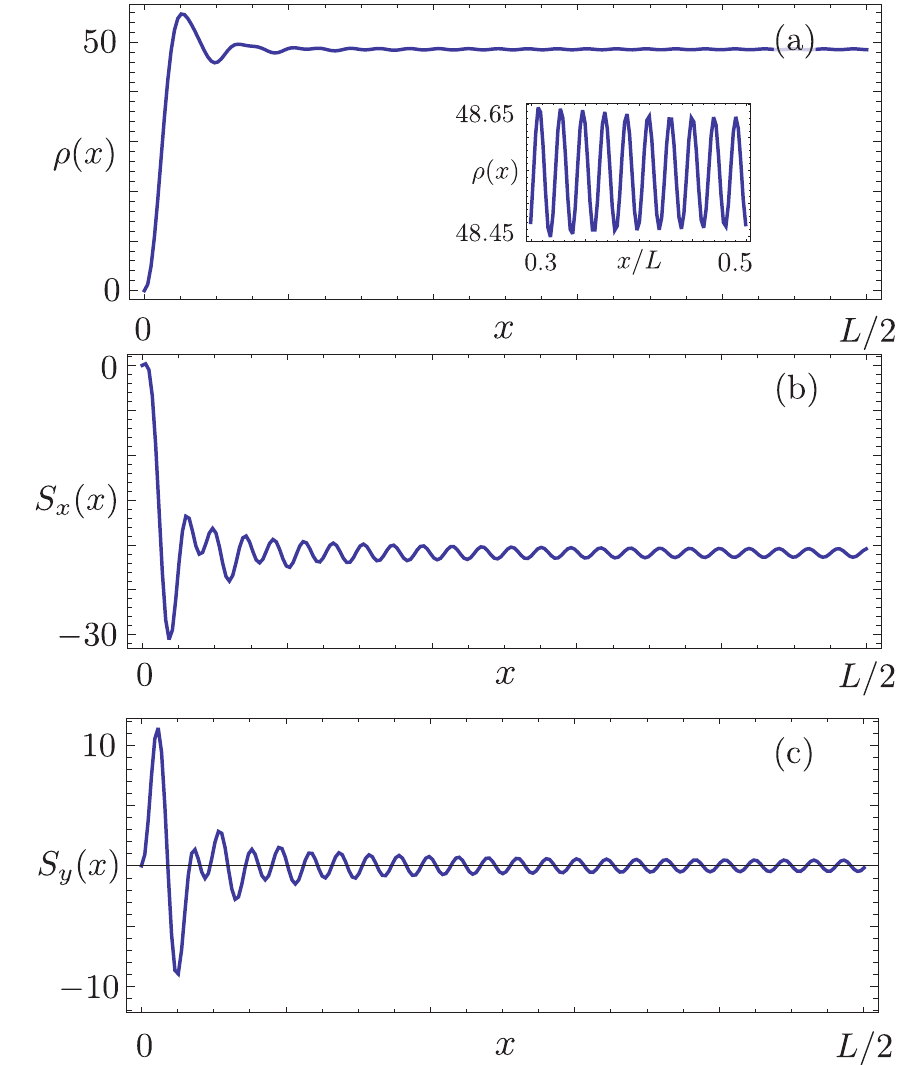}
\caption{Same as in Fig.~\ref{fig:fig1h} but for $\delta=0.177$. Here, $\beta=4000$ with other parameters as in Fig.~\ref{fig:fig1h}.}
\label{fig:fig2h}
\end{figure}
Increasing $\delta$, the ratio between the states in the gap and those below it, increases. However, as also discussed above, the $k_{F}$ oscillations induced by states in the gap tend to be confined only near the dot edges while the dot center becomes dominated by $2k_{F}$ oscillations. As a result, one would expect a mixed character of the charge density, with $2k_{F}$ oscillations in the center and $k_{F}$ oscillations near the dot edges. Analogously, one can expect a more regular pattern of the spin density oscillations. This is confirmed by the numerical results of the Hubbard model shown in Fig.~\ref{fig:fig2h} for $\delta=0.177$. The expected behavior is indeed observed.\\
\begin{figure}
\centering
\includegraphics[width=\columnwidth]{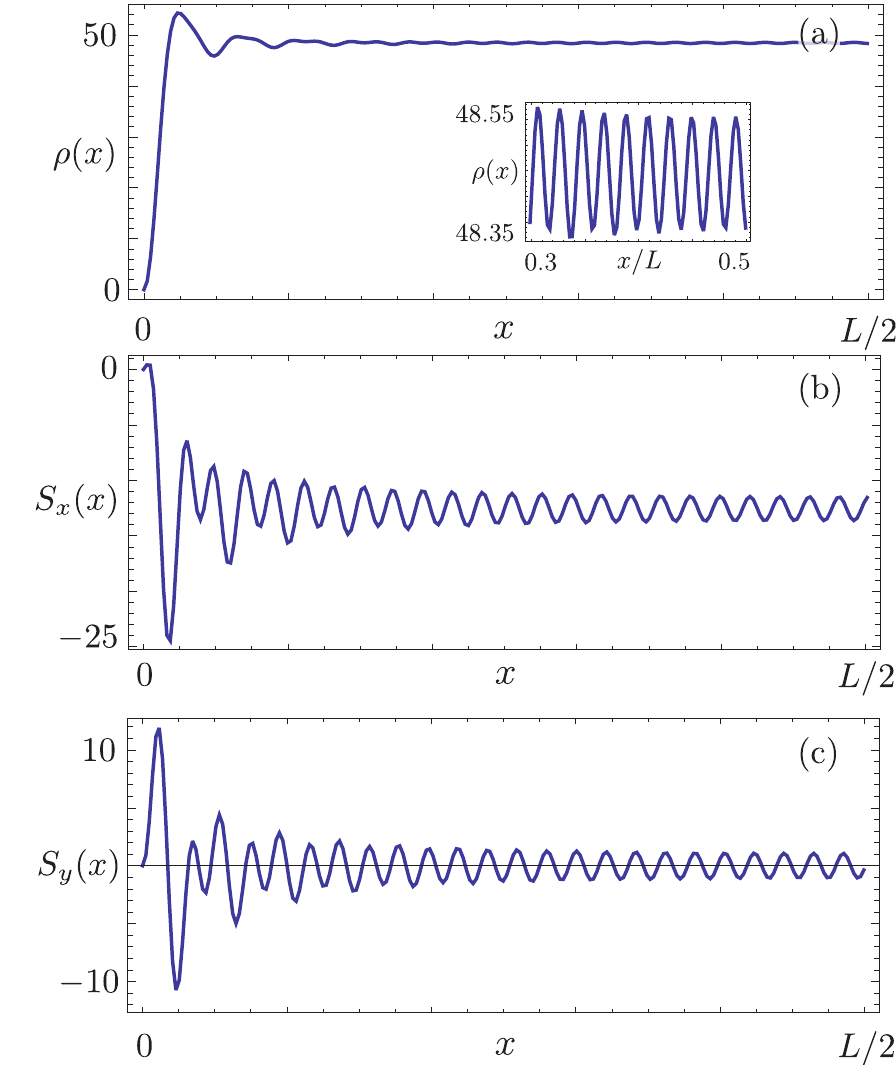}
\caption{Same as in Fig.~\ref{fig:fig1h} but for $U=t/2$. Other parameters as in Fig.~\ref{fig:fig1h}.}
\label{fig:fig3h}
\end{figure}
\noindent Let us now consider the effects of electron interactions. Figure~\ref{fig:fig3h} shows the situation for $\delta=0.044$ and $U=t/2$. Here, calculations are performed via a MPS variational algorithm~\cite{Essler:2005}. Oscillations of $\rho(x)$ with wavevector $2k_{F}$ are now evident in the dot center, while few ones with wavevector $k_{F}$ are present in a narrow region near the edges. We still attribute this effect to the role of the Q-H states near the Fermi surface in shaping the electron density. We also notice that the spin density exhibits more regular oscillations, as it occurs for the CEM when interactions are turned on. Both these facts support the idea that even in the interacting regime, features of the Q-H states near the Fermi surface can be detected in the charge and spin densities of the system.\\

\noindent We close this section with a brief comparison between the results provided by the Luttinger model and those of the Hubbard model. A common ground can be found in the {\em difference} between charge and spin densities for $N_{0}$ and $N_0-1$ electrons, $\delta\rho(x)=\left.\rho(x)\right|_{N_{0}}-\left.\rho(x)\right|_{N_{0}-1}$ and $\delta S_{\nu}(x)=\left.S_{\nu}(x)\right|_{N_{0}}-\left.S_{\nu}(x)\right|_{N_{0}-1}$ respectively. These quantities can be experimentally probed, e.g. in the shift of the chemical potential induced by a charged~\cite{Ziani:2012} or magnetized~\cite{Dolcetto:2013b} tip. In the non-interacting case, it is immediate to show that $\delta\rho(x)\equiv\varphi_{0}(x)$ and $\delta S_{\nu}(x)\equiv\Sigma_{0}^{(\nu)}(x)$. The results of the Luttinger and Hubbard model in this regime coincide within numerical accuracy (not shown). In the interacting case the density differences depend essentially on states at the Fermi surface within the limits of the Luttinger model developed here. 
\begin{figure}
\centering
\includegraphics[width=\columnwidth]{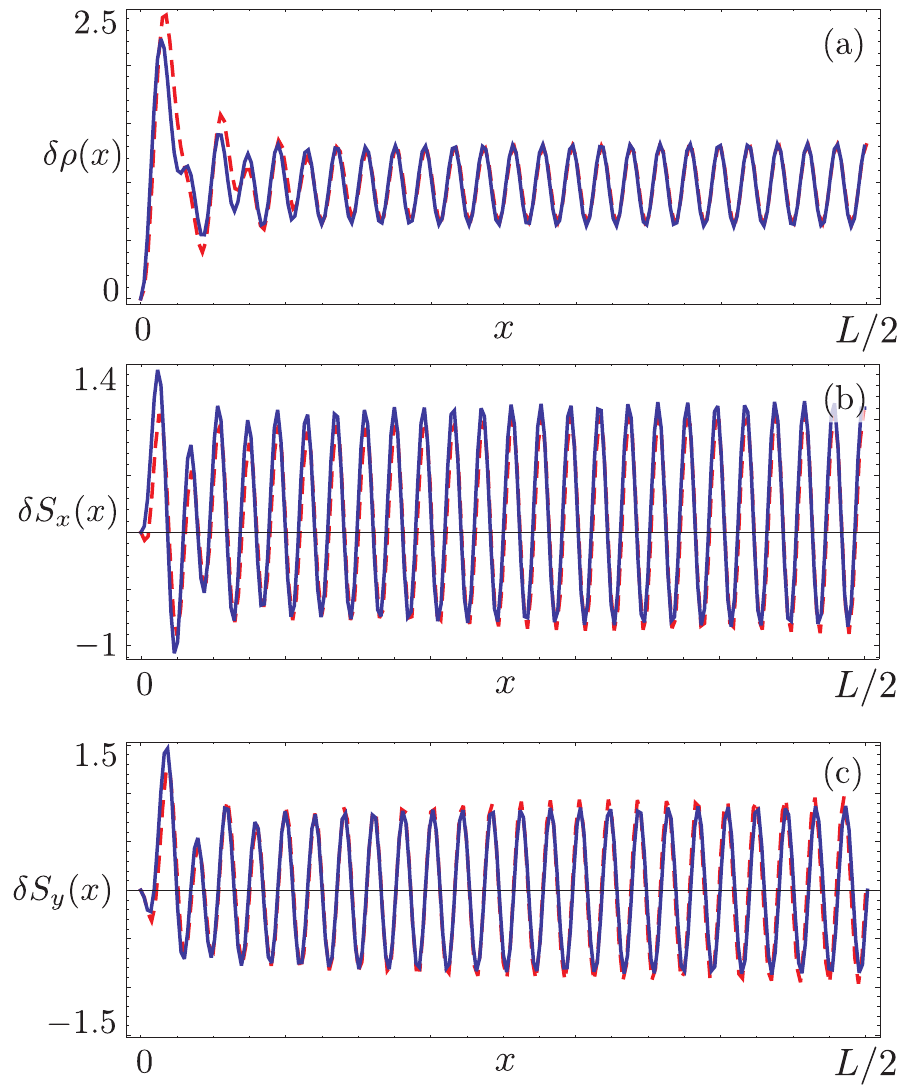}
\caption{(a) Plot of $\delta\rho(x)$ for $N_{0}=49$, $\delta=0.177$ and $U=t/2$ calculated with the Luttinger model (blue, solid) and the Hubbard model (red, dashed); (b) same as in (a) but for $\delta S_{x}(x)$; (c) same as in (a) but for $\delta S_{y}(x)$. Here, we have optimized the agreement between the curves by fitting the Luttinger parameter $g=0.78$. Other parameters as in Fig.~\ref{fig:fig3h}.}
\label{fig:fig4dmu}
\end{figure}
Figure~\ref{fig:fig4dmu} shows $\delta\rho(x)$ and $\delta S_{\nu}(x)$ for the case $U=t/2$, with $N_0=49$ and $\delta=0.177$. The Luttinger parameter has been fitted here to be $g\approx 0.8$ by maximizing the overlapping between the three calculated quantities. Such value is well within the range of validity of our model (see Eq.~\eqref{ll:eq:g_min}). As is clear, the agreement between the two models is indeed very good.

  
\section{Conclusions}\label{sec:conclusions}

We have studied the properties of a quantum dot embedded into a spin-orbit quantum wire subject to an external magnetic field. Focusing on the regime where Q-H states develop in the energy gap, we have built an analytical Luttinger theory with open boundaries. Both charge and spin properties of the Q-H states have been
analyzed, focusing on the collective excitation wavefunction and magnetization respectively. In the Q-H regime they respectively display {\em anomalous Friedel oscillations}, characterized by a wavevector $k_{F}$ instead of the expected $2k_{F}$, and peculiar spin textures in which the Q-H states magnetization precesses with a competition between oscillations at wavevectors $k_{F}$ at the edges and $2k_{F}$ around the center. Such effects are due to the presence of dot edge states, the hallmark of the Q-H regime, which tend to occupy the entire length of the dot in the Q-H regime and for weak interactions. When the magnetic field is increased, or interactions get stronger, the anomalous Friedel oscillations and peculiar spin textures tend to disappear from the dot center, while they are more robust near the edges. Signatures of these peculiar states can be detected also in the charge and spin densities although the effects on the latter quantities are much weaker.\\

\noindent The parameters employed in the paper can be achieved in state-of-the-art
systems. Indeed, for the case of InAs nanowires, the typical values of the parameters that appear in Eq.~\eqref{spm:qe:parameters} are~\cite{Sidor:2007,Csonka:2008,Liang:2012}: $\eta\approx2\cdot10^{-11}\ \mathrm{eV}\cdot\mathrm{m}$, $g^{*}\approx5$, $m^{*}\approx0.023\ m_{e}$, where $ m_{e} $ is the electron mass. To obtain the values $ \alpha=150 $, $ \beta=1000 $ and $ \beta=4000 $ used in the paper one has to impose $ L\approx 10\ \mathrm{\mu m} $ and, respectively, $ B=0.07\ \mathrm{T} $, $ B=0.28\ \mathrm{T} $. 


\begin{acknowledgements}
We acknowledge the financial support of the priority program SPP 1666 ”Topological insulators” (N.T.Z),  MIUR- PRIN-2010LLKJBX (S.B.) and MIUR-FIRB2012, Grant No. RBFR1236VV (F.M.G., F.C., M.S.). We are grateful to D. Rossini for kindly providing us with the MPS code. S.B. would like to thank L. Mazza for useful discussions.
\end{acknowledgements}


\appendix
\section{Secular equation}\label{sec:secular_equation}
In this Appendix we outline the solution of the secular equation in Eq.~\eqref{spm:eq:secular}
\begin{multline}
\alpha^2\sqrt{\beta^2-\varepsilon^2}\left[\cos(k_{1}L)\cosh(\kappa_{3}L)-1\right]=\\
(\alpha^2\varepsilon+2\beta^2)\sin(k_{1}L)\sinh(\kappa_{3}L).
\label{app:eq:secular}
\end{multline}
Since we are working in the limit $ |\varepsilon|\ll\beta $, Eq.~\eqref{spm:eq:k1k3} can be expanded to the first order in $ \varepsilon/\beta $ ($ k_3=i\kappa_3 $):
\begin{subequations}
\begin{align}
k_1&\approx k_1^{(0)}\left( 1+\eta_1\frac{\varepsilon}{\beta} \right), \label{app:eq:k1}\\
\kappa_3 &\approx \kappa_3^{(0)}\left( 1+\eta_3\frac{\varepsilon}{\beta} \right), 
\end{align}
\end{subequations}
with 
\begin{subequations}
\begin{align}
k_1^{(0)}&=k_1(\varepsilon=0)=\frac{\alpha}{L}\left(\frac{\sqrt{1+4\delta^2}+1}{2}\right)^{1/2},\\\kappa_3^{(0)}&=\kappa_3(\varepsilon=0)=\frac{\alpha}{L}\left(\frac{\sqrt{1+4\delta^2}-1}{2}\right)^{1/2},\label{app:eq:kappa3}
\end{align}
\end{subequations}
and
\begin{equation}
\eta_{1}=\frac{\delta}{\sqrt{1+4\delta^2}},\qquad \eta_{3}=-\eta_{1}.
\end{equation}
The critical condition to solve Eq.~\eqref{app:eq:secular} is the large number of states in the gap, $ k_{+}-k_{-}\gg\pi/L $, see Fig.~\ref{spm:fig:Bands}. For $ \delta\ll1/2 $ it casts into $ \beta/\alpha\gg1 $, while for $ \delta\gg1/2 $ it becomes $ \sqrt{\beta}\gg1 $. In both regimes it can be seen from Eq.~\eqref{app:eq:kappa3} that the latter are equivalent to impose $ \kappa_{3}^{(0)}L\gg1 $.  Thus, Eq.~\eqref{app:eq:secular} becomes 
\begin{equation}
\mathrm{cotan}(k_{1}L)=\frac{\alpha^2\varepsilon+2\beta^2}{ \alpha^2\sqrt{\beta^2-\varepsilon^2} }\approx\frac{\varepsilon}{\beta}+2\delta,
\end{equation}
that is
\begin{align}
k_{1}L&=\frac{\pi}{2}(2n+1)-\arctan\left(\frac{\varepsilon}{\beta}+2\delta\right)\nonumber\\
&\approx\frac{\pi}{2}(2n+1)-\frac{1}{1+4\delta^2}\frac{\varepsilon}{\beta}-\arctan(2\delta).
\end{align}
Substituting Eq.~\eqref{app:eq:k1} and solving for $ \varepsilon $, we obtain the discretized energy spectrum
\begin{equation}
\varepsilon_{n}=v_{0}\left[\frac{\pi}{2}(2n+1)-\arctan(2\delta)-k_{1}^{(0)}\right],
\label{app:eq:discspectrum}
\end{equation}
with
\begin{equation}
v_{0}=\beta/\eta_{1}k_{1}^{(0)}.
\end{equation}
If we now substitute the spectrum of Eq.~\eqref{app:eq:discspectrum} in Eq.~\eqref{app:eq:k1}, we also get the discretization of the wavevector: 
\begin{equation}
k_{1,n}\approx\frac{\pi}{2L}(2n+1)-\frac{1}{L}\arctan(2\delta).
\end{equation}
In the end, the spectrum assumes the simple form
\begin{equation}
\varepsilon_{k_{1,n}}=v_0\left[k_{1,n}-k_{1}^{(0)}\right].
\end{equation}
In a similar way one could obtain the discretization for $ \kappa_3 $ \emph{but}, since $ \kappa_3 $ appears only in exponential terms (see Eq.~\eqref{spm:eq:general_psi_kappa}) and $ |\eta_3 \varepsilon_n|\ll 1 $, we will ignore the corrections due to discretization and set $ \kappa_{3,n}\approx \kappa_{3}^{(0)} $ in all formulas.

\section{Interacting Hamiltonian} \label{sec:InteractingH}
In this Appendix we discuss the form of the interacting Hamiltonian of Eq.~\eqref{ll:eq:H_int_reduced}. We start from a general interacting Hamiltonian
\begin{equation}
\hat{H}_{\mathrm{int}}=\frac{1}{2}\int_{0}^{L}\int_{0}^{L}\hat{\Psi}^{\dagger}(x)\hat{\Psi}^{\dagger}(y)V(x-y)\hat{\Psi}(y)\hat{\Psi}(x)dxdy,
\label{appb:eq:H_int_gen}
\end{equation}
with a short range two-particle interaction $ V(x-y)$. Following the standard g-ology model~\cite{Solyom:1979}, we assign different coupling constants to different processes. Inserting in $ \hat{\Psi}(x) $ the expressions for $ \hat{\psi}_{\sigma}(x) $ in Eq.~\eqref{ll:eq:psi_upR}, $ \hat{H}_{\mathrm{int}} $ can be rewritten in term of $ \hat{\psi}_{R}(x) $. Terms that cannot be rewritten as product of two $ \hat{\rho}_{R} $ densities (Umklapp processes) are damped by fast oscillating exponentials $ \exp(\pm2ik_Fx) $, $ \exp(\pm4ik_Fx) $ and thus will be ignored. Introducing the notation $ A_{ij}=c^{*}_ic_j+d^{*}_id_j $ and using the relations $ A_{11}=1 $ and $ A_{22}=1 $, Eq.~\eqref{appb:eq:H_int_gen} becomes
\begin{equation}
\hat{H}_{\mathrm{int}}=\hat{H}_{\mathrm{int}}^{(0)}+\hat{H}_{\mathrm{int}}^{\mathrm{extra}}.
\end{equation}
Here
\begin{equation}
\hat{H}_{\mathrm{int}}^{(0)}=\frac{1}{2}\int_{-L}^{L}\left[g_4\hat{\rho}_R(x)\hat{\rho}_R(x)+\tilde{g}_2(\delta)\hat{\rho}_R(x)\hat{\rho}_{R}(-x)\right]dx,
\end{equation}
with
\begin{equation}
\tilde{g}_2(\delta)=g_2-g_1f(\delta),
\end{equation} 
where
\begin{equation}
f(\delta)=\left(\frac{2\delta}{1+\sqrt{1+4\delta^2}}\right)^2.
\label{appb:eq:fdelta}
\end{equation}
We now set~\cite{Voit:1995,Giamarchi:2004}, $ g_2=g_4=V(0) $ and $ g_1=V(2k_F)=\xi V(0) $, with $\xi \ll1 $,~\cite{Schulz:1993,Sassetti:1998,*Cuniberti:1996,Cavaliere:2004,*Dolcetto:2012} and define the parameter 
\begin{equation}
g=\left[1+\frac{V(0)}{\pi v_0}\right]^{-1/2}.
\end{equation}
All the terms $ \propto \hat{\rho}_R(c)$ with $ c\in\{0,L\} $ are included in $ \hat{H}_{\mathrm{int}}^{\mathrm{extra}} $. Here we can identify five classes of terms: 
\begin{subequations}
\begin{align}
&\propto\int_{0}^{L} e^{-2\kappa_3^{(0)}x}\hat{\rho}_R(0)\hat{\rho}_{R}(\pm x) dx, \label{appb:eq:rho0rhox}\\
&\propto \int_{0}^{L}e^{2\kappa_3^{(0)}(x-L)}\hat{\rho}_R(L)\hat{\rho}_{R}(\pm x)dx,\label{appb:eq:rhoLrhox}\\
&\propto \int_{0}^{L}e^{-4\kappa_3^{(0)}L}\hat{\rho}_R(0)\hat{\rho}_{R}(0)dx,\label{appb:eq:rho0rho0}\\
&\propto \int_{0}^{L}e^{-2\kappa_3^{(0)}L}\hat{\rho}_R(0)\hat{\rho}_{R}(L)dx,\label{appb:eq:rho0rhoL}\\
&\propto \int_{0}^{L}e^{4\kappa_3^{(0)}(x-L)}\hat{\rho}_R(L)\hat{\rho}_{R}(L)dx.\label{appb:eq:rhoLrhoL}
\end{align}
\end{subequations}
Note that, since $ \hat{\rho}_{R}(x) $ varies slowly with $ x $, the terms in Eqns.~(\ref{appb:eq:rho0rhox}, \ref{appb:eq:rhoLrhox}, \ref{appb:eq:rhoLrhoL}) are similar and can be estimated to be $ \propto 1/\kappa_{3}^{(0)}L $ while (\ref{appb:eq:rho0rho0}, \ref{appb:eq:rho0rhoL}) are $ \propto e^{-a\kappa_{3}^{(0)}L} $ (with $ a\in\{2,4\} $). Thus, in the limit $ \kappa_{3}^{(0)}L\gg1 $ one can approximate $ \hat{H}_{\mathrm{int}}\approx\hat{H}_{\mathrm{int}}^{(0)}  $. From Eq.~\eqref{ll:eq:rhoR}, the total Hamiltonian $ \hat{H}=\hat{H}_{0}+\hat{H}_{\mathrm{int}}^{(0)} $, being quadratic in the bosonic operator $ \hat{b}_{q} $, can be written in the diagonal form~\cite{Voit:1995,Giamarchi:2004}
\begin{equation}
\hat{H}=\bar{v}(\delta)\sum_{q>0}q\hat{d}_{q}^{\dagger}\hat{d}_{q}+\frac{\pi \bar{v}_{N}(\delta)}{2L}\hat{N}^{2}.
\end{equation}
Here the velocities of bosonic and zero mode are
\begin{subequations}
\begin{align}
\bar{v}(\delta)&=\frac{v_0}{K'(\delta)},\\
 \bar{v}_{N}(\delta)&=\frac{\bar{v}(\delta)}{K(\delta)},
\end{align}
\end{subequations}
where 
\begin{subequations}
\begin{align}
K'(\delta)&=\left\{\left(1+\frac{g_{4}}{2\pi v_{0}}\right)^{2}-\left[\frac{\tilde{g}_{2}(\delta)}{2\pi v_{0}}\right]^{2}\right\}^{-\frac{1}{2}},\\
\intertext{and}
K(\delta)&=\left[\frac{2\pi v_0+g_4-\tilde{g}_2(\delta)}{2\pi v_0+g_4+\bar{g}_2(\delta)}\right]^{\frac{1}{2}}. 
\end{align}
\label{appb:eq:KK}
\end{subequations}
We neglect here the weak intrinsic dependence of $ g $ on $ \delta $ due to $ v_0 $, which is particularly weak for $ \delta\lesssim1/2 $, the regime of interest in this paper. Equations~\eqref{appb:eq:KK} can be rewritten as
\begin{subequations}
\begin{align}
K'(\delta,g,\xi)\!&=\!\left\{g^{-2}\!+\!\left[\!\xi f(\delta)\!-\!\frac{1}{4}\xi^2f^2(\delta)\!\right]\!(g^{-2}\!-\!1)^{2} \right\}^{-\frac{1}{2}}\!\!\!,\\ K(\delta,g,\xi)&=\left[\frac{1+\frac{\xi}{2}(g^{-2}-1)f(\delta)}{g^{-2}-\frac{\xi}{2}(g^{-2}-1)f(\delta)}\right]^{\frac{1}{2}} ,
\end{align}
\label{appb:eq:KK2}
\end{subequations}
where $ f(\delta) $ is defined in Eq.~\eqref{appb:eq:fdelta}. For $ g=1 $ one has $ g=K(\delta,1,\xi)=K'(\delta,1,\xi)=1 $ $ \forall\,\delta,\,\xi $. Table~\ref{appb:tab} shows the comparison between $ g $, $ K(\delta,g,\xi) $ and $ K'((\delta,g,\xi)) $ for~\cite{Schulz:1993} $ \xi=0.1 $  and for different values of $ \delta $. The typical values of $ g $ chosen here fall within the region of validity of the theory we develop (see Eq.~\eqref{ll:eq:g_min}). One can immediately see that $ g\approx K(\delta,g,\xi)  \approx K'(\delta,g,\xi) $. This means that the contribution of the term $ \propto g_1=V(2k_F) $ is negligible. Thus we can set $ K(\delta,g,\xi) =K'(\delta,g,\xi) =g $ in all the above relations and observe that they turn into the ones that follow from the simplified interacting Hamiltonian adopted in the main text (see Eq.~\eqref{ll:eq:H_int_reduced}). 
 \begin{table}
 \centering
 \caption{Comparison between the values of $ g $, $ K(\delta,g,\xi) $ and $ K'(\delta,g,\xi) $ for fixed values of $ \delta $ and $ \xi $. Here $ \xi=0.1 $.   }
 \begin{ruledtabular}
 \begin{tabular}{lccc}

$ g  $                  &$ \delta $ & $ K(\delta,g,\xi) $ & $ K'(\delta,g,\xi)  $ \\
\hline
                                 
\multirow{3}*{$ 0.8 $} & $ 0.05 $ & $ 0.800046 $ & $ 0.799980 $ \\
							       & $ 0.2 $   & $ 0.800684 $ & $ 0.799700 $ \\
                                   & $ 0.5 $   &$ 0.803169 $ & $ 0.798620 $ \\
                                   & $ 2 $      &$ 0.811292 $ & $ 0.795181 $ \\
\hline                                   
\multirow{3}*{$ 0.6 $} & $ 0.05 $& $ 0.600090 $ & $ 0.599915 $ \\
							       & $ 0.2 $& $ 0.601345 $ & $ 0.598739 $ \\
                                   & $ 0.5 $& $ 0.606224 $ & $ 0.594252 $ \\                
                                  & $ 2 $ & $ 0.622140 $ & $ 0.580502 $ \\                        
                                   
 \end{tabular}
 \end{ruledtabular}
 \label{appb:tab}
 \end{table}

\section{Tunneling through an STM tip}
\label{sec:appstm}
Spatial oscillations of states near the Fermi surface can be probed experimentally by  tunneling of electrons in the linear transport regime\cite{Bruus_Flensberg:2004} via a (possibly, magnetized) STM tip~\cite{Ziani:2011, Ziani:2013}. Applying a suitable bias to the tip it is  for instance possible to inject electrons with a given spin direction (which needs not to coincide with the quantization axis) at a specific location $x$ of the dot. Electrons then tunnel through the barriers at the edges and flow to drain contacts. In the tunneling limit, when the tunneling through the tip is the slowest process, it can be shown that the linear conductance is essentially determined by the tunneling rate through the tip only~\cite{Ziani:2011, Ziani:2013}. From now on, we will therefore focus on this rate solely. The tunneling coupling between the tip and the dot is described by the Hamiltonian
\begin{equation}
\hat{H}_{T}^{(\nu)}=\tau \sum_{k,\sigma}\hat{\psi}_{\sigma}^{\dagger}(x)\hat{c}_{k,\sigma}^{(\nu)}	
\end{equation}
where $\tau$ is the tunneling amplitude, $\sigma=\uparrow,\downarrow$ the spin direction (referred to the $ z $ axis), $k$ the wavevector in the tip and $\nu$ the quantization axis of the spin {\em in the tip}. Furthermore, $\hat{c}_{k,\sigma}^{(\nu)}$ are operators for electrons in the tip, represented on the eigenbasis of $\sigma_{z}$. They are connected to the operators on the natural basis of spin eigenstates along the direction $\nu$, $\hat{d}_{p}^{(\nu)}$ with $p=+$ ($p=-$) for electrons with spin parallel (antiparallel) to the $\nu$ axis, by 
\begin{eqnarray}
\hat{c}_{k,\uparrow}^{(x)}&=&\frac{1}{\sqrt{2}}\left(-\hat{d}_{k,+}^{(\nu)}+\hat{d}_{k,-}^{(\nu)}\right),\\	
\hat{c}_{k,\downarrow}^{(x)}&=&\frac{1}{\sqrt{2}}\left(\hat{d}_{k,+}^{(\nu)}+\hat{d}_{k,-}^{(\nu)}\right),\\	
\hat{c}_{k,\uparrow}^{(y)}&=&\frac{1}{\sqrt{2}}\left(\hat{d}_{k,+}^{(\nu)}+i\hat{d}_{k,-}^{(\nu)}\right),\\
\hat{c}_{k,\downarrow}^{(y)}&=&\frac{i}{\sqrt{2}}\left(\hat{d}_{k,+}^{(\nu)}-i\hat{d}_{k,-}^{(\nu)}\right) .
\end{eqnarray}
Consider an electron with spin orientation $p$ along the axis $\nu$ tunneling into a dot with $N_{0}$ electrons at position $x$. The tunneling rate for such a process is given by~\cite{Bruus_Flensberg:2004, Ziani:2011, Ziani:2013}
\begin{eqnarray}
\Gamma^{(\nu)}_{p}(x)&=&2\pi\mathcal{D}_{p}^{(\nu)}\left[1-f(\Delta E_{D}+\Delta E_{T})\right]\nonumber\\
&&\left|\langle N_0+1|\hat{O}_{(p)}^{(\nu)}(x)|N_0\rangle\right|^2\label{eq:trat}
\end{eqnarray}
where
\begin{eqnarray}
\hat{O}_{(+)}^{(x)}(x)&=&\frac{1}{\sqrt{2}}\left[\hat{\psi}_{\downarrow}^{\dagger}(x)-\hat{\psi}_{\uparrow}^{\dagger}(x)\right]\, ,\label{eq:O1}\\
\hat{O}_{(-)}^{(x)}(x)&=&\frac{1}{\sqrt{2}}\left[\hat{\psi}_{\downarrow}^{\dagger}(x)+\hat{\psi}_{\uparrow}^{\dagger}(x)\right]\, ,\label{eq:O2}\\
\hat{O}_{(+)}^{(y)}(x)&=&\frac{1}{\sqrt{2}}\left[\hat{\psi}_{\uparrow}^{\dagger}(x)+i\hat{\psi}_{\downarrow}^{\dagger}(x)\right]\, ,\label{eq:O3}\\
\hat{O}_{(-)}^{(y)}(x)&=&\frac{i}{\sqrt{2}}\left[\hat{\psi}_{\uparrow}^{\dagger}(x)-i\hat{\psi}_{\downarrow}^{\dagger}(x)\right]\, .\label{eq:O4}
\end{eqnarray}
Furthermore, $\mathcal{D}_{p}^{(\nu)}$ is the density of states of electrons in the tip with spin orientation $p$ along the $\nu$ direction, $f(E)$ is a Fermi function and $\Delta E_{D,T}$ are the energy differences between final and initial dot and tip states respectively. By looking up the definitions in Eqns.~(\ref{eq:O1}-\ref{eq:O4}) it is immediately clear that Eqns.~(\ref{eq:ratetot},\ref{eq:ratesp}) hold.

\bibliography{GambettaBibliography.bib}

\end{document}